\documentclass[aps,showpacs,prd,twocolumn,nofootinbib,nobibnotes]{revtex4-2}
\usepackage{graphicx}
\usepackage{amssymb}
\usepackage{amsmath}
\usepackage{color}
\usepackage{float}
\usepackage{ulem}
\usepackage{accents}
\usepackage{graphicx}
\usepackage{graphicx}
\usepackage{amsfonts}
\usepackage[colorlinks=true,
pdfstartview=FitV,linkcolor=blue,
citecolor=blue,urlcolor=blue,breaklinks=true]{hyperref}
\usepackage{array}
\usepackage{float}
\usepackage{placeins}
\usepackage[dvipsnames]{xcolor}
\usepackage{csquotes}
\usepackage{bbold}
\usepackage{units}
\usepackage{fixmath}
\usepackage{cancel}

\newcolumntype{C}[1]{>{\centering\arraybackslash}m{#1}}
\renewcommand{\eqref}[1]{\mbox{Eq.~(\ref{#1})}}

\definecolor{ForestForestGreen}{rgb}{0.13,0.55,0.13}

\definecolor{ForestGreen}{rgb}{0.13,0.55,0.13}

\begin{document}

\title{\Large A Drude-Lorentz dielectric in the presence of a magnetic current density}

\author{Pedro D. S. Silva$^{a}$}
\email{pedro.dss@discente.ufma.br} \email{pdiego.10@hotmail.com}
\author{M. J. Neves$^{b}$}\email{mariojr@ufrrj.br}
\author{Manoel M. Ferreira Jr.$^{a,c}$}
\email{manoel.messias@ufma.br; manojr.ufma@gmail.com}
\affiliation{$^a$Programa de P\'{o}s-graduaç\~{a}o em F\'{i}sica, Universidade Federal do Maranh\~{a}o, Campus Universit\'{a}rio do Bacanga, S\~{a}o Lu\'is (MA), 65080-805, Brazil}
\affiliation{$^{b}$Departamento de F\'{i}sica, Universidade Federal Rural do Rio de Janeiro,
BR 465-07, Serop\'edica (RJ), 23890-971, Brazil}
\affiliation{$^c$Departamento de F\'{\i}sica, Universidade Federal do Maranh\~{a}o,
Campus Universit\'{a}rio do Bacanga, S\~{a}o Lu\'is (MA), 65080-805, Brazil}

\date{\today}

\begin{abstract}

The dispersive propagation and absorption in the domain of a Drude-Lorentz dielectric modified by a magnetic current are classically investigated, taking as the starting point the Drude electric permittivity, in connection with the chiral magnetic effect. Using standard electromagnetic methods, the dispersion relations and refractive indices are evaluated and discussed, as well as the group and the energy velocities. We consider three configurations for the magnetic conductivity tensor: isotropic, antisymmetric, and symmetric. In all these cases, we carry out the group velocity and the energy velocity, for investigating the properties of signal propagation. We observe that the isotropic magnetic conductivity enhances the signal propagation at low and intermediary frequencies, while the symmetric and antisymmetric conductivity severely constrains the wave propagation in these frequency windows. 

\end{abstract}



\maketitle

\pagestyle{myheadings}
\markright{A Drude-Lorentz dielectric in the presence of a magnetic current density}

\section{Introduction}
The Maxwell's electrodynamics a general framework to describe the electromagnetic interaction
in both classical and quantum scales. The classical electrodynamics explains the interaction of charges in
motion via solutions of Maxwell's equations \cite{Zangwill,Jackson,Landau}. The quantum electrodynamics (QED) describes with excellent precision the interaction of the photons with leptons at the MeV-scale \cite{FeynmanQED}.
The electrodynamics applied to the material physics is one of the basic tools of Condensed Matter Physics
(CMP). In theoretical CMP, many models are developed to study the electronic properties of solids. One of the examples
is the Drude model, initially used to explain the properties of the electrons transported through the metals \cite{Drude1,Drude2}, providing a complex and dispersive conductivity and refractive index associated with the absorption of electromagnetic waves in metals \cite{Ascrofit}. The Drude model is also applied to describe the
optical propagation in dielectric media \cite{Ascrofit,Fox,Wooten}, in which the electromagnetic waves
interact with vibrating atoms, considered as classical dipoles. This kind of approach allows us to obtain optical properties of solids, including dispersion, absorption and optical activity.
In some specific continuous media, there appears a magnetic current density linear in the magnetic field,  originated from an asymmetry between the density of the left- and right-handed chiral fermions \cite{Kharzeev1,Kharzeev2,Kharzeev3}.
This is known as a macroscopical manifestation of the quantum Chiral Magnetic Effect (CME), being investigated in different research areas in physics, as Cosmology, quark-gluon plasm, electroweak interactions and others \cite{Dvornikov,Schober,Dvornikov2,Dvornikov3,Dvornikov4}. In CMP, the CME is associated with the Weyl semimetals (WSM) in which massless fermions acquire velocity along the magnetic field whose direction is driven by the chirality of these fermions \cite{Burkov}. In WSM, the magnetic current density may present a general form, ${\bf{J}}=\alpha \, (\bf{E} \cdot \bf{B}) \, {\bf{B}}$ \cite{Li,Xiaochun-Huang,Barnes}, with $\alpha$ being a constant, in such a way the magnetic current,
\begin{eqnarray}
\label{J}
{\bf{J}}=\sigma^{B} \cdot {\bf{B}} \; ,
\end{eqnarray}
is written in terms of an anisotropic magnetic conductivity tensor denoted by $\sigma^{B}_{ij}$, which depends on the material properties. Classical optical properties of a dielectric medium endowed magnetic current were already investigated, supposing the medium is described by usual constitutive relations, $\mathbf{D}={\epsilon }\,\mathbf{E}$ and $\mathbf{B}={\mu }\,\mathbf{H}$ \cite{Pedro1}. The refractive indices, propagating modes, and optical effects were examined for three configurations of the magnetic conductivity tensor,
\begin{equation}
{\sigma}_{ij}^{B}=\Sigma \, \delta_{ij} \; , \quad 
{\sigma}_{ij}^{B}={\epsilon}_{ijk} \, b_{k} \; , \quad
\sigma_{ij}^{B}=\frac{1}{2}\left(a_{i}\,c_{j}+a_{j}\,c_{i}\right) \; ,
\label{eq68-1}
\end{equation}
associated with isotropic, antisymmetric ~\cite{Kaushik1} and symmetric currents, given as
\begin{align}\label{JS}
{\bf{J}}=\Sigma \, {\bf{B}} \;, \quad 	{\bf{J}}={\bf{b}}\times {\bf{B}} \;,
\quad {\bf{J}}=\frac{1}{2} \left[(\bf{a}\cdot \bf{B})\bf{c}+(\bf{c}\cdot \bf{B})\bf{a} \right] \; ,
\end{align}
where $\bf{a}$, $\bf{b}$, $\bf{c}$ are constant vectors in space.

%



%

%
The electrodynamics of a continuous medium is described by the Lagrangian density,
\begin{eqnarray}\label{LEDmatter}
{\cal L}=-\frac{1}{4} \, G_{\mu\nu}F^{\mu\nu}-J_{\mu}A^{\mu} \; ,
\end{eqnarray}
where $F^{\mu\nu}=\partial_{\mu}A_{\nu}-\partial_{\nu}A_{\mu}$ and
$G^{\mu\nu}=\chi^{\mu\nu\rho\lambda}F_{\rho\lambda}/2$ are the field strength electromagnetic tensors in vacuum and matter, and $J^{\mu}=(\rho,{\bf J})$
is the source four-current\footnote{We adopt natural units $\hbar=c=1$,
and the Minkowski metric $\eta^{\mu\nu}=\mbox{diag}(+1,-1,-1,-1)$ throughout this paper.} The constitive tensor
$\chi^{\mu\nu\rho\lambda}$ satisfies the symmetry properties
$\chi^{\mu\nu\rho\lambda}=-\chi^{\nu\mu\rho\lambda}$, $\chi^{\mu\nu\rho\lambda}=-\chi^{\mu\nu\lambda\rho}$
and $\chi^{\mu\nu\rho\lambda}=\chi^{\rho\lambda\mu\nu}$. The action principle
applied to the Lagrangian (\ref{LEDmatter}) yields the field equation
\begin{eqnarray}\label{eqG}
\partial_{\mu}G^{\mu\nu}=J^{\nu} \; ,
\end{eqnarray}
while the EM strength field tensor satisfies the Bianchi identity,
$\partial_{\mu}\widetilde{F}^{\mu\nu}=0$, where $\widetilde{F}^{\mu\nu}=\epsilon^{\mu\nu\rho\lambda}F_{\rho\lambda}/2$
is the dual tensor. Explicitly, the motion equations can be written in the standard form
\begin{subequations}
\begin{eqnarray}
\nabla\cdot{\bf D} \!&=&\! 0
\hspace{0.3cm} , \hspace{0.3cm}
\nabla\times{\bf E}+\frac{\partial{\bf B}}{\partial t} = {\bf 0} \; ,
\label{eqdivD}
\\
\nabla\cdot{\bf B} \!&=&\!0
\hspace{0.3cm} , \hspace{0.3cm}
\nabla\times{\bf H}-\frac{\partial{\bf D}}{\partial t} = {\bf J} \; ,
\label{eqdivB}
\end{eqnarray}
\end{subequations}
where ${\bf D}$ and ${\bf H}$ are the fields in the matter represented
by the components $G^{0i}=-D^{i}$ and $H^{i}=\epsilon^{ijk}G^{jk}/2$.
In this paper, we use the constitutive relations,
\begin{align}
{\bf{D}} &= \epsilon (\omega) \, {\bf{E}} \; , \quad {\bf{H}} =\frac{1}{\mu} \, {\bf{B}} \; , \label{constitutive-relations1}
\end{align}
applied to a Drude-Lorentz dielectric, whose complex permittivity
\begin{align}
\epsilon(\omega)=\epsilon_{0}\left[ 1+\frac{\omega_{p}^2}{\omega_{0}^2-\omega^2-i\,\gamma\,\omega} \right] \; ,
\label{epsilonomega}
\end{align}
takes into account the electron's damped oscillation. Here, $\omega_{0}$ is the free electron frequency, $\gamma$ is the damping factor and $\omega_{p}=\sqrt{ nq^2/\epsilon_{0}m }$ is the plasma frequency.  From the Drude-Lorentz model, $q=-e$ is the electron charge,
$m$ is the electron mass, and $n$ is the number density. Equivalently, one can write \eqref{epsilonomega} as
\label{drude-permittivity-definition-1}
\begin{align}
 \epsilon(\omega) &= \epsilon' (\omega) + i \epsilon''(\omega) \; , \label{epsilon-simplification-1}
\end{align}
where
\begin{subequations}
\label{general-Drude-Lorentz-epsilon-0}
\begin{align}
\epsilon' (\omega) &= \epsilon_{0} \left[1 + \frac{\omega_{p}^{2} \, (\omega_{0}^{2} -\omega^{2})}{ (\omega_{0}^{2} - \omega^{2})^{2} + \gamma^{2} \omega^{2}} \right] \; , \label{real-part-drude-epsilon-definition} \\
\epsilon'' (\omega) &= \epsilon_{0} \left[ \frac{ \gamma \, \omega_{p}^{2} \, \omega} { (\omega_{0}^{2} - \omega^{2})^{2} + \gamma^{2} \omega^{2} } \right] \; .
\label{imaginary-part-drude-epsilon-definition}
\end{align}
\end{subequations}

This Drude dielectric, described by the usual constitutive relations (\ref{constitutive-relations1}), will be examined in the presence of the non conventional current constitutive  relation, 
\begin{equation}
	{J}_{i}={\sigma}_{ij}^{B} \, {B}_{j} \; .
	\label{JCMEij}
\end{equation}
Substituting the plane wave solutions, ${\bf E}({\bf r},t)={\bf E} \, e^{i({\bf k}\cdot{\bf r}-\omega t)}$
and ${\bf B}({\bf r},t)={\bf B} \, e^{i({\bf k}\cdot{\bf r}-\omega t)}$ in the Maxwell relations, and using the current density (\ref{JCMEij}), we obtain the following equation for the electric field ${\bf E}$:
\begin{align}
	\label{EqwaveE0}
	\left[ \, {\bf k}^2 \, \delta_{ij} - k_{i}\,k_{j}-\omega^2 \, \mu \, \overline{\epsilon}_{ij}(\omega) \, \right] E^{j}=0 \; ,
\end{align}
with the effective permittivity,
\begin{eqnarray}\label{barepsilonij}
	\overline{\epsilon}_{ij}(\omega)= \epsilon(\omega) \, \delta_{ij}
	-\frac{i}{\omega^2} \, \sigma^{B}_{ia}\epsilon_{abj}\,k_{b} \;,
\end{eqnarray}
where $\sigma^{B}_{ij}$ is the  magnetic conductivity tensor, and $\epsilon(\omega)$ is the dielectric permittivity. Equation (\ref{EqwaveE0}) can be written in terms of the refractive index components $n^{i}=k^{i}/\omega$, in which the refractive index of the medium is defined by $n=+\sqrt{n_{i} \, n_{i}}$.
Thereby, (\ref{EqwaveE0}) can be cast into the form
\begin{align}
	M_{ij}E^{j}=0 \; , \label{Fresnel-equation-1}
\end{align}
in which the tensor $M_{ij}$ reads
\begin{eqnarray}
	M_{ij}=n^2 \, \delta_{ij}- n_{i} \, n_{j} - \mu \, \overline{\epsilon}_{ij}(\omega) \; . \label{Fresnel-equation-2}
\end{eqnarray}
The nontrivial solutions for $E^{j}$ in \eqref{Fresnel-equation-1} are obtained by requiring $\mathrm{det}[M_{ij}]=0$, which provides the dispersion relations of the model and all the useful information to obtain the group and energy velocities.
In dispersive media the electromagnetic fields may be complex,
${\bf E}={\bf E}^{\prime} + i \, {\bf E}^{\prime\prime}$, ${\bf B}={\bf B}^{\prime} + i \, {\bf B}^{\prime\prime}$. Additionally, in the presence of absorption the wave vector is complex,
	\begin{equation}
		{\bf k}={\bf k}^{\prime} + i \, {\bf k}^{\prime\prime} \; ,
	\end{equation}
with the energy balance being ruled by the Poynting's theorem \cite{Zangwill}, whose Poynting vector and electromagnetic energy density (stored in the field) are
\begin{subequations}
	\begin{eqnarray}
			{\bf S}\!&=&\left({\bf E}\times{\bf H}^{\ast}\right)=\frac{1}{2\mu'}\left({\bf E}\times{\bf B}^{\ast}\right) \; ,	\label{complex-group-velocity-isotropic-28}
		\\
		u_{EM} \!&=&\! \frac{1}{2} \frac{\partial [\omega \epsilon' ]}{\partial \omega} \left( {\bf{E}} \cdot {\bf{E}}^{*} \right) + \frac{1}{2\mu'} \left( {\bf{B}} \cdot {\bf{B}}^{*} \right) \;.
	\end{eqnarray}
\end{subequations}
In the absence of sources $(\rho=0)$, the  Gauss's law is compatible with the attainment of transversal modes,
\begin{subequations}
\begin{align}
	\label{complex-group-velocity-isotropic-54A}
	({\bf{k}}' \cdot {\bf{E}}') = ({\bf{k}}'' \cdot {\bf{E}}'')=0 \; , \\
	({\bf{k}}' \cdot {\bf{E}}'') =({\bf{k}}'' \cdot {\bf{E}}')=0 \; , \label{complex-group-velocity-isotropic-54B}
\end{align}
\end{subequations}
for which the time-averaged quantities read
\begin{subequations}
	\begin{align}\label{timeaverageSuEM1}
		\left\langle {\bf{S}} \right\rangle &=  \frac{{\bf{E}}^{2}}{2\omega\mu'} \, {\bf k}^{\prime}   \; ,
		\\
		\left\langle u_{EM} \right\rangle &= \frac{1}{4} \left[ \epsilon' + \omega \, \frac{ \partial \epsilon'}{\partial \omega} + \frac{ |{\bf{k}}|^{2}}{\mu \omega^{2}} \right] {\bf{E}}^{2} \;.
		\label{timeaverageSuEM2}
	\end{align}
\end{subequations}
In lossy media, the ratio of these quantities defines the energy velocity,
\begin{align}
	\label{energy-velocity-26}
	{\bf{v}}_{E} &= \frac{ \left\langle {\bf{S}} \right\rangle} {\left\langle u_{EM} \right \rangle} \; ,
\end{align}
suitable to describe the energy transport in the presence of absorption.

In a non-absorbing dieletric ($\epsilon^{\prime\prime}=0$), the group velocity
\begin{align}\label{complex-group-velocity-isotropic-26}
	{\bf{v}}_{g} &= \frac{d \omega}{d k} \, \hat{\bf{k}}  \; ,
\end{align}
is a real quantity suitable for representing the signal propagation. However, in an absorbing medium, the group velocity becomes complex, as 	
${\bf{v}}_{g} = \mathrm{Re}[d\omega/dk] + i \,  \mathrm{Im}[d\omega/dk]$, no longer describing energy transport, being replaced by the energy velocity \cite{Loudon,Loudon2,Sherman,Davidovich,Ruppin,Nunes} in this respect. 

For the Drude dielectric permittivity, Eq. (\ref{timeaverageSuEM2}) is rewritten as
 		\begin{align}
 			\langle u_{EM} \rangle = \frac{\epsilon_{0} |{\bf{E}}|^{2} }{4} \left[ 1 + \frac{\omega_{p}^{2} (\omega_{0}^{2} -\omega^{2})}{ \Gamma_{+}(\omega)} + \frac{2\omega^2 \omega_{p}^{2} \Gamma_{-}}{\Gamma_{+}^{2}(\omega)} + \frac{|{\bf k}|^{2}}{\mu \epsilon_{0} \omega^{2}} \right],
 			\label{energy-momentum-tensor-complex-4}
 		\end{align}
 with $\Gamma_{\pm}(\omega)=(\omega_{0}^{2} - \omega^{2})^{2} \pm\gamma^{2} \omega^{2}$, in such a way the energy velocity reads
\begin{align}
\label{VEresult}
{\bf v}_{E}= \frac{ {\bf k}^{\prime} }{\mu\epsilon_0 \omega}\left[1 + \frac{\omega_{p}^{2} (\omega_{0}^{2} -\omega^{2})}{ \Gamma_{+}(\omega)} + \frac{2\omega^2 \omega_{p}^{2} \Gamma_{-}}{\Gamma_{+}^{2}(\omega)} + \frac{|{\bf k}|^{2}}{\mu \epsilon_{0} \omega^{2}} \right]^{-1},
\end{align}
By definition, ${\bf v}_{E}$ points along the $\hat{{\bf k}}^{\prime}$-direction, also depending on the dispersion relation of $k(\omega)$. In general non absorbing dielectrics, energy velocity and group velocity are the same one. But such an equality that is lost in lossy media, where it holds ${\bf v}_{E} \neq {\bf v}_{g}$. Such results have been recently verified for dispersive, absorbing and lossless hyperbolic metamaterial (HMM) \cite{Moradi}, as far as general bi-anisotropic media \cite{Darinskii}. In accordance with the definition (\ref{energy-velocity-26}), the energy velocity depends on the way the energy density and the dissipated energy density in a lossy medium are written \cite{Cui,Semchenko}, which constitute an active line of research actually  \cite{Chen21,Chen22}. 
An alternative procedure to investigate group and energy velocities in dissipative dynamical systems \cite{Gerasik} was used recently to investigate the group and energy propagation in a dielectric medium endowed with magnetic conductivity \cite{PedroMario}. Such a study was performed considering  $\epsilon^{\prime}=cte$ and $\epsilon^{\prime\prime}=0$ or $\epsilon^{\prime}=cte$ and $\epsilon^{\prime \prime}=\sigma \omega^{-1}$, comparing group velocity and energy velocity for the cases of isotropic, symmetric and antisymmetric magnetic conductivity shown in Eq. (\ref{eq68-1}).
 In this paper, we investigate the properties of propagation and absorption in a realistic Drude dielectric (DD) endowed with magnetic conductivity following the procedures of Ref. \cite{PedroMario}. Initially, we obtain the solutions of refractive indices, the propagating modes for the scenarios of an isotropic conductivity magnetic, including evaluation of the group and the energy velocities. Posteriorly, it is done for the case of anisotropic conductivity magnetic tensor, antisymmetric and a symmetric tensor. The features of the refractive indices are discussed having as comparison basis the DD ones. The paper is organized as follows. Section \ref{section-isotropic-case} is dedicated to the case the isotropic tensor $\sigma^{B}_{ij}$. In Section \ref{section-antisymmetric-case} we obtain the results for an antisymmetric magnetic conductivity tensor parametrized by one constant vector. Section \ref{section-symmetric-case} is devoted to examinining the symmetric conductivity tensor configuration parametrized by two constant vectors. All these sections are divided in two subsections in which the optical properties and the group/energy velocities aspects are addressed. For end, in the section \ref{final}, we present the conclusions and the final remarks.

\section{\label{section-isotropic-case}The case of the isotropic magnetic conductivity tensor}
The magnetic current density with a diagonal current, the isotropic conductivity magnetic tensor is
\begin{eqnarray}
\sigma^{B}_{ij}=\Sigma \, \delta_{ij} \; ,
\end{eqnarray}
where $\Sigma$ is a real and positive constant, in which the bar permittivity tensor is
\begin{eqnarray}
\overline{\epsilon}_{ij}(\omega)= \epsilon(\omega) \, \delta_{ij} +\frac{i\,\Sigma}{\omega} \, \epsilon_{ijk} \, n_{k} \; .
\end{eqnarray}

The null determinant of the $M_{ij}$ matrix implies into the $n$-polynomial equation
\begin{eqnarray}\label{eqndiagonal}
\left[ \, n^2 -\mu\,\epsilon(\omega) \, \right]^2-n^2 \, \frac{\mu^2\Sigma^2}{\omega^2}=0 \; .
\end{eqnarray}
The solutions of the equation (\ref{eqndiagonal}) are given by
\begin{align}
\label{indices-isotropic-case-1}
n_{\pm} &= \sqrt{\mu(\epsilon' (\omega) + i \epsilon''(\omega)) + \left( \frac{\mu \Sigma}{2\omega} \right)^{2} } \pm \frac{\mu \Sigma}{2\omega} \; .
\end{align}

\begin{figure}[H]
	\includegraphics[width=8.42cm]{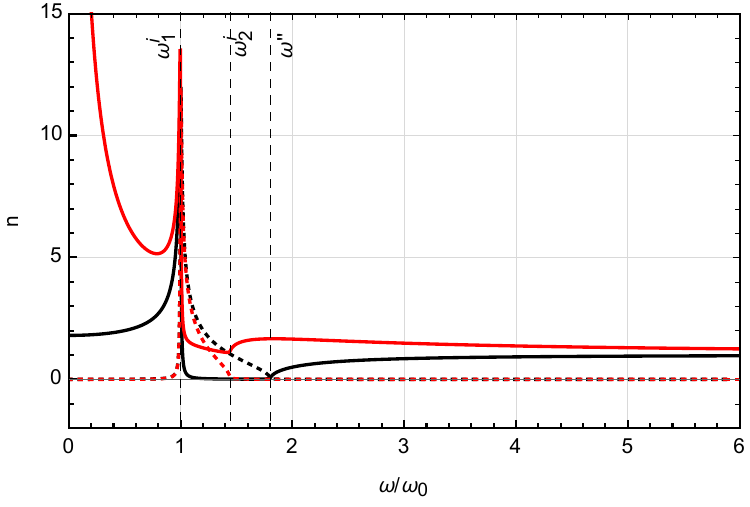}
	\caption{Refractive index $n_{+}$ of \eqref{indices-real-imaginary-partes-general-1}. Solid (dotted) curves represent the real (imaginary) pieces of $n_{+}$. The black lines represent the usual case, where $\Sigma=0$. The red ones represent the index for $\Sigma\neq0$. Here, we have used: $\mu=1$, $\epsilon_{0}=1$, $\omega_{p} /\omega_{0}=1.5$, $\gamma/\omega_{0}=0.01$, and $\Sigma/\omega_{0}=3$. The frequency $\omega_{2}^{i}$ marks the end of absorption zone, which is now shortened by the presence of the magnetic conductivity.} 
	\label{plot-n-mais-isotropic-case-Drude-with-magnetic-current}
\end{figure}

\begin{figure}[h]
\includegraphics[width=8.42cm]{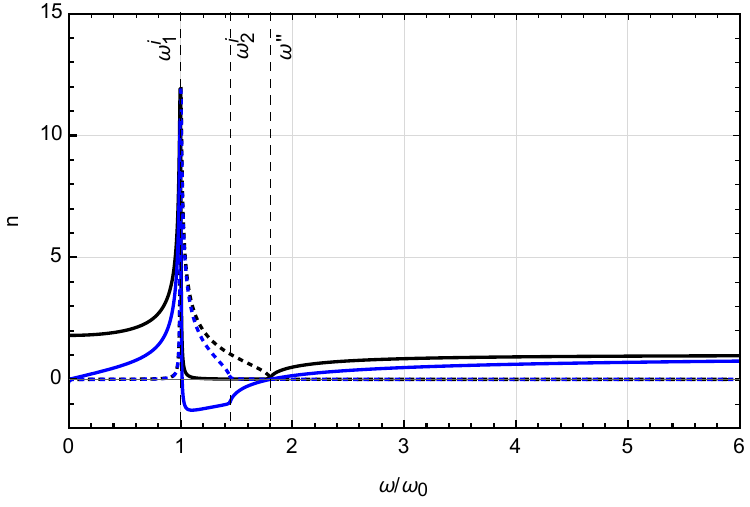}
\caption{Refractive index $n_{-}$ of \eqref{indices-real-imaginary-partes-general-1}. Solid (dotted) curves represent the real (imaginary) pieces of $n_{-}$. Black curves indicate the usual case, where $\Sigma=0$. The blue ones represent the index for $\Sigma\neq0$. Here, we have used: $\mu=1$, $\epsilon_{0}=1$, $\omega_{p} /\omega_{0}=1.5$, $\gamma/\omega_{0}=0.01$, and $\Sigma/\omega_{0}=3$. The absorption window narrowing becomes evident for $\Sigma\neq0$. }
\label{plot-n-menos-isotropic-case-Drude-with-magnetic-current}
\end{figure}

Substituting the dielectric permittivity (\ref{general-Drude-Lorentz-epsilon-0}), both solutions can be split into their real and imaginary pieces,
\begin{eqnarray}
n_{\pm} &=\Re[n_{\pm}] + i \, \Im [n_{\pm}] \; , \label{indices-real-imaginary-partes-general-1}
\end{eqnarray}
where
\begin{subequations}
\begin{align}
\Re[n_{\pm}] &=  \sqrt{\frac{|h(\omega)|}{2} } \, I_{+} \pm \frac{\mu \Sigma}{2\omega} \; , \label{real-n-isotropic-case-1}\\
\Im[n_{\pm}] &=  \sqrt{\frac{|h(\omega)|}{2} } \, I_{-} \; , \label{imaginary-n-isotropic-1}
\end{align}
\end{subequations}
with
\begin{subequations}
\begin{align}\label{Imaismenos}
I_{\pm} &\equiv \sqrt{ \sqrt{ 1 + \mu^2 \, \left[\frac{ \epsilon''(\omega)}{h(\omega)} \right]^{2} } \pm \mbox{sgn}[h(\omega)] } \; ,
\\
h(\omega) &=\mu\epsilon^{\prime}(\omega)+\left(\frac{\mu\Sigma}{2\omega}\right)^2 \; ,
\label{A-functions-definitions-1}
\end{align}
\end{subequations}
where $\mbox{sgn}$ represents the signal function of $h(\omega)$.
Notice that $\epsilon^{\prime}(\omega)$ from (\ref{real-part-drude-epsilon-definition}) can assume negative values for a
given range of $\omega$-frequency. Consequently, the $h(\omega)$-function can also be negative.
The general behavior of $n_{\pm}$ real and imaginary pieces, written in Eqs. (\ref{real-n-isotropic-case-1}) and (\ref{imaginary-n-isotropic-1}), is illustrated in Figs.~\ref{plot-n-mais-isotropic-case-Drude-with-magnetic-current} and \ref{plot-n-menos-isotropic-case-Drude-with-magnetic-current} in terms of the dimensionless variable $\omega / \omega_{0}$. We note that both $\Re[n_{\pm}]$ indicate propagation regime for the range $\omega < \omega_{0}$. Near the origin the real piece of $n_{+}$ tends to infinity, an analog behavior of magnetized plasmas \cite{Filipe}. At the resonance frequency, $\omega=\omega_{0}$, there occurs an absorption peak which coincides with the beginning of the window for $\Im[n_{\pm}]\neq 0$. Furthermore, immediately after $\omega=\omega_{0}$ the values of $\Re[n_{\pm}]$ rapidly decrease (anomalous dispersion), being followed by a short increasing region  for the index $n_{+}$, which in the sequel exhibits another anomalous region (extending to the asymptotic limit); see Fig. \ref{plot-n-mais-isotropic-case-Drude-with-magnetic-current}. As for the index $n_{-}$, however, the anomalous dispersion region is followed by a normal dispersion one; see Fig. \ref{plot-n-menos-isotropic-case-Drude-with-magnetic-current}. For both indices, the absorption peak occurs in the region $\omega_{1}^{i} < \omega < \omega_{2}^{i}$, with $\omega_{1}^{i} \approx \omega_{0}$, which defines the absorption window. Note that near the origin, the real piece of $n_{-}$ is null, which is distinct from the conventional DD behavior.

It is worth mentioning that negative refraction occurs for $n_{-}$ in the zone $\omega_{1}^{i} <\omega < \omega''$, which also contains the absorption frequency window. This negative refraction is a consequence of the isotropic magnetic conductivity. For $n_{\pm}$, the frequency $\omega_{2}^{i}$ defines the absorption zone limit, which is reduced in the presence of the magnetic conductivity. It is also interesting to note that, besides the absorption width reduction, the indices $n_{\pm}$ also develop a non-null real piece, enabling energy transport in this region, not allowed in the correspondent usual DD one.  For $\omega >\omega_{2}^{i}$, the absorption is null and attenuation-free propagation occurs. For very high-frequencies $(\omega \gg\omega_{0})$, one obtains $\Re[n_{\pm}] \rightarrow \sqrt{\mu\epsilon_{0}}$.

For very low-frequencies ($\omega \ll \omega_{0}$), the refractive indices behave as
\begin{align}
n_{\pm} \simeq \sqrt{\mu \epsilon_{0} \left( 1 + \frac{\omega_{p}^{2}}{\omega_{0}^{2}} \right) + \left(\frac{\mu \Sigma}{2\omega} \right)^{2}} \pm \frac{ \mu \Sigma} {2\omega} \; , \label{indices-isotropic-case-low-frequencies}
\end{align}
which near the origin yields $n_{-} = 0$ and $n_{+} \simeq \mu \Sigma / \omega$, thus explaining the $n_{+}$ divergence at $\omega\rightarrow 0$ in Fig. \ref{plot-n-mais-isotropic-case-Drude-with-magnetic-current}. It is worth to point out that such low-frequency behavior, $\Re[n_{+}]\rightarrow \infty $, $\Re[n_{-}]\rightarrow 0$, is already present in the case of a simple dielectric with constant magnetic conductivity $\epsilon^{\prime}=cte$, $\epsilon^{\prime\prime}=\sigma/\omega$, $\Sigma=cte$. See Ref. \cite{PedroMario}. As a consistency check, one notes that when $\Sigma \rightarrow 0$, the refractive indices of \eqref{indices-isotropic-case-low-frequencies} yield a finite and non-null value in the origin, recovering the standard scenario. See the black continuous lines in Figs. \ref{plot-n-mais-isotropic-case-Drude-with-magnetic-current} and \ref{plot-n-menos-isotropic-case-Drude-with-magnetic-current}.  

The peak of the imaginary part is located at $\omega \approx \omega_{0}$. Using \eqref{imaginary-n-isotropic-1} and expressions (\ref{real-part-drude-epsilon-definition}), (\ref{imaginary-part-drude-epsilon-definition}), the peak amplitude is
\begin{align}
\left. \frac{}{} \Im(n_{\pm})\right|_{\omega=\omega_{0}} &= \sqrt{ \frac{\mu \epsilon_{0}}{2} + \left( \frac{ \mu \Sigma}{2\sqrt{2}\omega_{0}} \right)^{2}} \, \mathcal{I} \; ,  \label{imaginary-part-indices-at-resonance-1}
\end{align}
with
\begin{align}
\mathcal{I} &=\sqrt{ \sqrt{ 1 + \left[ \frac{ 4 \omega_{0} \omega_{p}^{2} \epsilon_{0}}{\gamma \left( 4\omega_{0}^{2} \epsilon_{0} + \mu \Sigma^{2} \right) } \right]^{2} } -1 } \; , \label{imaginary-part-indices-at-resonance-2}
\end{align}
revealing that the magnetic conductivity also modifies the absorption peak magnitude. In the next section, we will discuss the consequence of $\Sigma$ on the imaginary part of $n_{\pm}$.

\subsection{Propagation modes and optical effects}
\begin{figure}[H]
\begin{centering}
\includegraphics[scale=0.68]{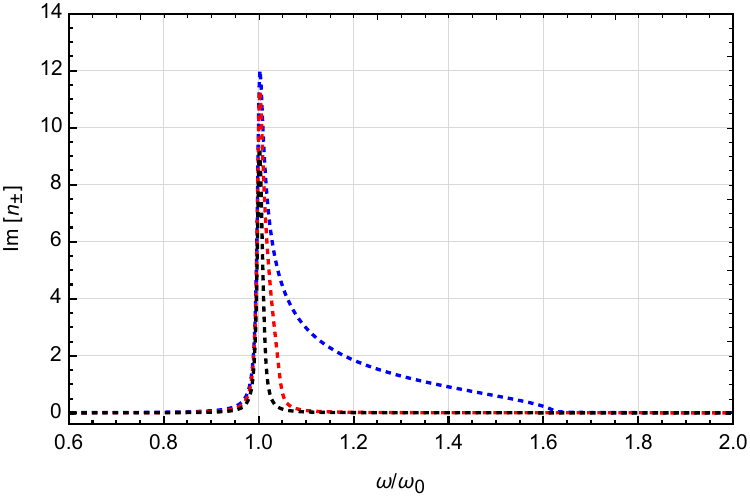}
\par\end{centering}
\caption{ Imaginary part of $n_{\pm}$ of \eqref{indices-real-imaginary-partes-general-1} in terms of $\omega/\omega_{0}$. Here, we have set $\mu=1$, $\epsilon_{0}=1$, $\omega_{p}/\omega_{0}=1.5$, $\gamma/\omega_{0}=0.01$, and $\Sigma/\omega_{0}=2$ (blue), $\Sigma/\omega_{0}=10$ (red), and $\Sigma/\omega_{0}=20$ (black).  }
\label{imaginary-part-isotropic-case-figure-2-version-2}
\end{figure}

Let us consider propagation along the ${\cal Z}$-axis, ${\bf{n}}= (0, 0, n)$, such that the tensor $M_{ij}$ becomes
\begin{align}
[M_{ij} ]&=
\left[
\begin{array}{ccc}
n^{2} - \mu \epsilon (\omega) & + i \mu \frac{\Sigma}{\omega} n & 0
\\
\\
-i\mu  \frac{\Sigma}{\omega} n & n^{2} - \mu \epsilon(\omega) & 0
\\
\\
0 & 0 & - \mu\epsilon(\omega)
\end{array}
\right] \; .
\label{propagating-modes-isotropic-1}
\end{align}
In order to find the propagating modes, we solve the equation $M_{ij}E^{j}=0$, including in the matrix (\ref{propagating-modes-isotropic-1}) the solution of \eqref{eqndiagonal}, that is,
\begin{align}
n^{2} - \mu\epsilon(\omega) = \pm \, n \, \frac{\mu \Sigma}{\omega} \; ,
\label{propagating-modes-isotropic-2}
\end{align}
providing the following electric fields:
\begin{align}
{\bf{E}}_{\pm} &= \frac{1}{\sqrt{2}} \begin{pmatrix}
1\\
\pm \, i\\
0
\end{pmatrix} \; , \label{propagating-modes-isotropic-3}
\end{align}
which represent left-handed (LCP) and right-handed (RCP) circularly polarized vectors, respectively. The two different real refractive indices associated with the modes (\ref{propagating-modes-isotropic-3}) imply circular birefringence, ascribed to the isotropic magnetic conductivity, $\Sigma$, and expressed in terms of the rotatory power (RP),
\begin{align}
\delta &= -\frac{\omega}{2} \left[ \Re(n_{+})-\Re(n_{-}) \right] \; , \label{rotatory-power-1}
\end{align}
specialized for the indices (\ref{indices-real-imaginary-partes-general-1}) as
\begin{align}
\delta &= -\frac{\mu \Sigma}{2} \; . \label{rotatory-power-isotropic-case-1}
\end{align}
It is a nondispersive RP that depends linearly on the magnetic conductivity, being equal to the one that appears in nondispersive dielectrics ($\epsilon^{\prime}=cte$) endowed with isotropic magnetic conductivity \cite{Pedro1}. 
It is interesting to mention that the absorption peak amplitude decreases with an increasing magnitude of the magnetic conductivity $\Sigma$.  This is in agreement with \eqref{imaginary-part-indices-at-resonance-1}, which provides $\Im(n_{\pm}) \rightarrow 0$ in the limit of very large conductivity $(\Sigma \rightarrow \infty)$.  In this limit, therefore, this Drude-Lorentz-like dielectric exhibits a nearly null absorption. Such an effect is represented by the different peak amplitudes in Fig.~\ref{imaginary-part-isotropic-case-figure-2-version-2}.

In lossy systems, the dichroism takes place when one propagating mode is more strongly absorbed than the other, which is directly associated to the imaginary piece of the refractive indices.  Dichroism represents a tool to distinguish between Dirac and Weyl semimetals \cite{Hosur}, being used also for enantiomeric discrimination \cite{Nieto-Vesperinas,Tang} and for developing graphene-based devices at terahertz frequencies \cite{Amin}. For circularly polarized modes, the difference of absorption between left and right-handed circularly polarized modes ~\cite{Shibata} is given by the circular dichroism coefficient,
\begin{equation}
	\label{eq:dichroism-power1}
	{\delta}_{\mathrm{d}}=-\frac{\omega}{2}[\mathrm{Im}(n_{+})-\mathrm{Im}(n_{-})]\,.
\end{equation}
Concerning the modes (\ref{propagating-modes-isotropic-3}) associated with the refractive indices (\ref{indices-real-imaginary-partes-general-1}), the dichroism coefficient is null, 
\begin{equation}
	\label{eq:dichroism-power1b}
	{\delta}_{\mathrm{d}}=0,
\end{equation}
since these indices have equal imaginary parts, see Eq. (\ref{imaginary-n-isotropic-1}). This is the same result obtained for the case of a simple dielectric with constant magnetic conductivity $\epsilon^{\prime}=cte$, $\Sigma=cte$ \cite{PedroMario}. In this sense, the Drude permittivity, although being complex and involved, does not modify either the RP or the dichroism coefficient in relation to the optical panorama of a simpler dielectric.

\subsection{The phase, group, and the energy velocities}
Implementing $n= \sqrt{{\bf{k}}^{2}} / \omega$ in \eqref{propagating-modes-isotropic-2}, we obtain
\begin{align}
k^{2} \mp \mu\Sigma \, k - \mu\,\omega^{2}\,\epsilon(\omega) = 0 \; ,
\label{group-velocity-isotropic-case-1}
\end{align}
whose the solutions read 
\begin{align}
k_{\pm} &= \omega \, \sqrt{ \mu \epsilon^{\prime}(\omega)+i\mu\epsilon^{\prime\prime}(\omega) + \frac{ \mu^{2} \Sigma^{2}} {4\omega^{2}}} \pm \frac{\mu \Sigma}{2} \; , \label{kvector2A}
\end{align}
with $\epsilon'$ and $\epsilon''$ of Eqs. (\ref{real-part-drude-epsilon-definition}) and (\ref{imaginary-part-drude-epsilon-definition}), respectively. These solutions can also be written as $k_{\pm}=k_{\pm}^{\prime}+i\,k_{\pm}^{\prime\prime}$, where
\begin{subequations}
\label{isotropic-case-real-and-imaginary-parts-of-momentum}
\begin{eqnarray}
k_{\pm}^{\prime}&=&\omega\,\sqrt{\frac{|h(\omega)|}{2} } \, I_{+} \pm \frac{\mu \Sigma}{2} \; ,
\label{RealkIsotropicpart}
\\
k_{\pm}^{\prime\prime}&=& \omega \, \sqrt{\frac{|h(\omega)|}{2} } \, I_{-}
\; ,
\label{ImkIsotropicpart}
\end{eqnarray}
\end{subequations}
and the functions $I_{\pm}$ and $h(\omega)$ are given in Eqs. (\ref{Imaismenos}) and
(\ref{A-functions-definitions-1}).

For the propagation analysis, we consider the real part \eqref{RealkIsotropicpart}, which represents the two distinct modes, $k_{+}$ and $k_{-}$.
The phase velocities are
\begin{align}
v_{\mathrm{ph}}^{(\pm)} &= \frac{\omega}{k^{\prime}_{\pm}} = \frac{1}{\Re[n_{\pm}]}
\; , \label{group-velocity-isotropic-case-4}
\end{align}
yielding the phase velocity difference,
$\Delta v_{\mathrm{ph}}=v_{\mathrm{ph}}^{(-)}-v_{\mathrm{ph}}^{(+)}$, namely,
\begin{align}
\Delta v_{\mathrm{ph}} &= \frac{4\omega\,\mu\,\Sigma}{2\omega^2\,|h(\omega)|\,I_{+}^2-\mu^2\,\Sigma^2 }
\; , \label{group-velocity-isotropic-case-5}
\end{align}
indicating that the magnetic conductivity $\Sigma$ engenders distinct modes propagating with different phase velocities and birefringence.
The group velocity is given in this case by
\begin{align}
v_{g}^{(\pm)} &=\frac{1}{\mu} \frac{ k_{\pm}/\omega \mp \mu \Sigma / (2 \omega)} {\epsilon(\omega)+(\partial_{\omega}\epsilon)\omega/2 }
\; ,  \label{group-velocity-isotropic-case-6}
\end{align}
with $\partial_{\omega}\epsilon=\partial \epsilon / \partial \omega$. Considering the expression (\ref{kvector2A}), the group velocity takes on the form,
	\begin{align}
		v_{g}^{\pm} = \frac{1}{\mu} \frac{ \displaystyle \sqrt{\mu \epsilon^{\prime}(\omega)+i\mu\epsilon^{\prime\prime}(\omega) + \frac{\mu^{2} \Sigma^{2}}{4 \omega^{2}}}}{ \displaystyle  \epsilon^{\prime}(\omega)+i\epsilon^{\prime\prime}(\omega) + \frac{\omega}{2}(\partial_{\omega}{\epsilon})}, \label{observation-10}
\end{align}
which can be rewritten by separating its real and imaginary pieces as follows:
\begin{subequations}
\begin{eqnarray}
\Re[v_g^{(\pm)}] \!&=&\! \frac{(\mu\epsilon_0)^{-1}}{C^2+D^2}
\sqrt{\frac{| h(\omega)|}{2}} \left( \, I_{+}\,C + \, I_{-} \, D \right) \; , \;\;
\label{group-velocity-isotropic-case-7Re}
\\
\Im[v_g^{(\pm)}] \!&=&\! \frac{(\mu\epsilon_0)^{-1}}{C^2+D^2}\sqrt{\frac{|h(\omega)|}{2}}
\left( \, I_{-}  \, C - I_{+} \, D \, \right) \; ,
\label{group-velocity-isotropic-case-7Im}
\end{eqnarray}
\end{subequations}
%
where
\begin{subequations}
\label{C-D-definitions}
\begin{align}
C &= \epsilon^{\prime}(\omega)+ \frac{\omega}{2}  \frac{\partial \epsilon^{\prime}}{\partial\omega} \,  ,
\label{group-velocity-isotropic-case-8}\\
D &= \epsilon^{\prime\prime}(\omega)+ \frac{\omega}{2}  \frac{\partial \epsilon^{\prime\prime}}{\partial\omega}  \; . \label{group-velocity-isotropic-case-9}
\end{align}
\end{subequations}

\begin{figure}[H]
\begin{centering}
\includegraphics[scale=0.68]{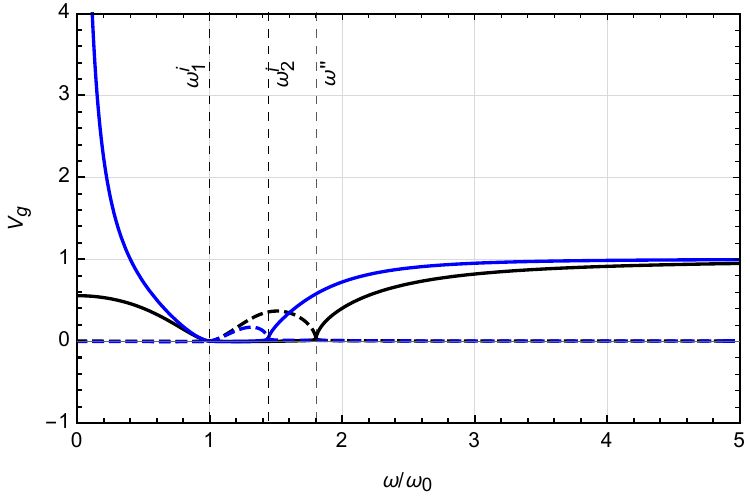}
\par\end{centering}
\caption{\label{group-velocity-isotropic-case} Group velocity of \eqref{group-velocity-isotropic-case-6} or \eqref{observation-10} as function of the dimensionless parameter $\omega/\omega_{0}$. Solid (dashed) lines represent the real (imaginary) parts of $v_{g}$, respectively. Here, we have used: $\mu=1$, $\epsilon_{0}=1$, $\omega_{p} /\omega_{0}=1.5$, $\gamma/\omega_{0}=0.01$, and $\Sigma/\omega_{0}=3$ (blue), and $\Sigma=0$ (black, usual case). The vertical dashed lines indicate $\omega/\omega_{0} \approx \{1, 1.44, 1.8\}$, from left to right, accordingly with \eqref{frequencies-isotropic-case}.}
\end{figure}

For the real part of the group velocity, see \eqref{group-velocity-isotropic-case-7Re}, it goes to infinity for very low-frequencies $(\omega \rightarrow 0)$, losing physical meaning. Figure \ref{group-velocity-isotropic-case} illustrates the behavior of $v_{g}$. In the high-frequency regime $(\omega \rightarrow \infty)$, one finds $\Re[v_g^{(\pm)}] \rightarrow 1/\sqrt{\mu\epsilon_0}$ and $\Im[v_g^{(\pm)}] \rightarrow 0$. This asymptotic behavior is the same one attained in the limit $\Sigma \rightarrow 0$ and $\omega_{p} \rightarrow 0$, which implies $\epsilon^{\prime} \rightarrow 1$ and $\epsilon^{\prime\prime} \rightarrow 0$, recovering a non-absorbing dielectric.

In scenarios where the group velocity is complex, the sign propagation is better represented by the energy velocity, $V_{E}$, which is here obtained by substituting the solutions (\ref{RealkIsotropicpart}) and (\ref{ImkIsotropicpart}) in
expression (\ref{VEresult}), yielding
\begin{subequations}
\label{energy-velocity-isotropic-7-0}
\begin{align}
V_{E}^{\pm} &= \frac{  \displaystyle  \sqrt{ 2  |h(\omega)|} \omega I_{+} \pm \mu\Sigma} {2 \omega U^{I}_{\pm}} \; , \label{energy-velocity-isotropic-7}
\end{align}
associated with $k_{+}^{\prime}$ and $k_{-}^{\prime}$, respectively, with
\begin{align}
U^{I}_{\pm}&= \frac{ \mu\epsilon'}{2} + \frac{\mu\omega}{2} \frac{\partial \epsilon'}{\partial \omega} + \frac{ |h(\omega)|}{4} (I_{+}^{2}+I_{-}^{2}) + \frac{ \mu^{2}\Sigma^{2}}{8\omega^{2}} + \nonumber \\
&\phantom{=}\pm \frac{\mu\Sigma}{2\omega} \, \sqrt{  \frac{ |h(\omega)|}{2}} \, I_{-} \; . \label{energy-velocity-isotropic-7-1}
\end{align}
\end{subequations}

\begin{figure}[h]
\begin{centering}
\includegraphics[scale=0.68]{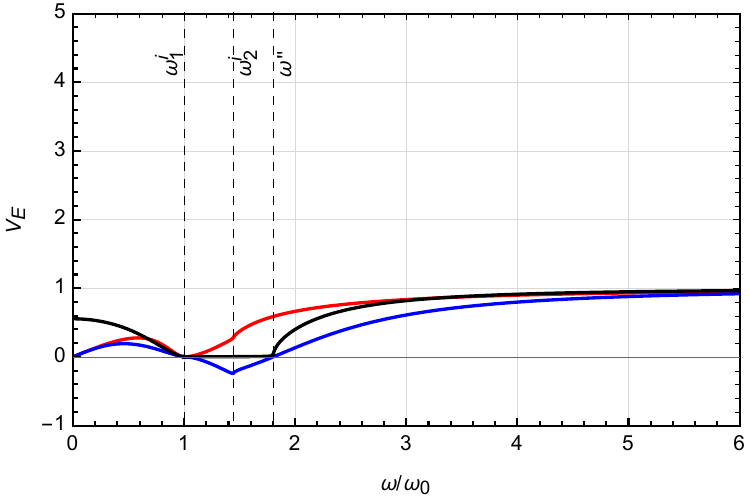}
\par\end{centering}
\caption{\label{plot-energy-velocity-isotropic-case-Drude-with-magnetic-current} Energy velocities of \eqref{energy-velocity-isotropic-7} in terms of the dimensionless parameter $\omega/\omega_{0}$. Red (blue) lines represent $V_{E}^{\pm}$, respectively. The black dashed curve illustrates the usual case (\ref{VEresult}), where $\Sigma=0$. Here, we have used: $\mu=1$, $\epsilon_{0}=1$, $\omega_{p} /\omega_{0}=1.5$, $\gamma/\omega_{0}=0.01$, and $\Sigma/\omega_{0}=3$. The vertical dashed lines indicate $\omega/\omega_{0} \approx \{1, 1.44, 1.8\}$, from left to right, accordingly with \eqref{frequencies-isotropic-case}.}
\end{figure}

Figure \ref{plot-energy-velocity-isotropic-case-Drude-with-magnetic-current} illustrates the energy velocities as function of the dimensionless parameter $\omega/\omega_{0}$. The characteristic frequencies $\omega^{i}_{1,2}$ are given by
the real non-negative solutions of the sixth-order equation in $\omega$,
\begin{eqnarray}
&&
\gamma^{2} \left( \omega^{4} + \frac{\mu \Sigma^{2}}{4\epsilon_{0}}  \omega_{0}^{2} \right) +(\omega_{0}^{2}-\omega^{2}) \times
\nonumber \\
&&
\times  \left[ \omega^{2} \left(\omega_{0}^{2}-\omega^{2}+\omega_{p}^{2} \right) + \frac{\mu\Sigma^{2}}{4\epsilon_{0}} \left(\omega_{0}^{2}-\omega^{2}-\gamma^{2} \right) \right] = 0 \; ,
\label{frequencies-isotropic-case}
\nonumber \\
\end{eqnarray}
and $\omega^{i}_{1,2}$ define the limits of the absorption zone for the propagating modes.
This absorption effect modifies the behavior of $V_{E}^{\pm}$ with the frequency, as one can see. The frequency $\omega''$ marks the end of the absorption window for a usual DD, where the energy velocity is vanishing. Note, however, that the energy velocity becomes non-null in the absorption zone of the DD modified by the magnetic conductivity since the refractive indices develop a finite real part. It implies the reduction of the lossy window, favoring the signal propagation.
	
It is also relevant to note the much more involved energy velocity of Fig. \ref{plot-energy-velocity-isotropic-case-Drude-with-magnetic-current} in relation to the correspondent one of Ref.~\cite{PedroMario}, obviously a consequence of the intricate Drude permittivity here considered.
The real non-negative solutions of \eqref{frequencies-isotropic-case} yield the characteristic frequencies $\omega^{i}_{1,2}$ when $\Sigma \neq 0$, and also the frequencies $\omega'$, $\omega''$ for $\Sigma = 0$. In the numerical examples we have considered, one obtains $\omega' \approx \omega^{i}_{1}$. This is the reason only $\omega^{i}_{1}$ appears in the plots.
Furthermore, we observe that:
\begin{itemize}
\item For $0<\omega <\omega^{i}_{1}$, both modes propagates without attenuation. In this regime, the energy velocity $V_{E}^{\pm}$ increases slowly with frequency, starting from zero. It decreases as long as it approaches the resonance frequency, $\omega \rightarrow \omega_{0} \approx \omega^{i}_{1}$, where there is absorption peaks indicated by the dotted lines in Figs.~\ref{plot-n-mais-isotropic-case-Drude-with-magnetic-current} and \ref{plot-n-menos-isotropic-case-Drude-with-magnetic-current}. At this frequency, the energy velocities $V_{E}^{+}$ and $V_{E}^{-}$ tend to zero. As for $V_{E}^{-}$, it exactly holds $V_{E}^{-}(\omega^{i}_{1})=0$.

\item For $\omega^{i}_{1}<\omega < \omega^{i}_{2}$, there occurs propagation and absorption simultaneously (non-null real and imaginary pieces of $n_{\pm}$), since the energy velocities assume non-null values. In fact, $V_{E}^{+}$ is positive in this interval (see the red curve in Fig.~\ref{plot-energy-velocity-isotropic-case-Drude-with-magnetic-current}. On the other hand, the velocity $V_{E}^{-}$ is negative in this zone, indicating negative refraction. This latter feature is in accordance with Fig.~\ref{plot-n-menos-isotropic-case-Drude-with-magnetic-current}, where the real piece of $n_{-}$ (blue solid curve) is negative in $\omega^{i}_{1}<\omega <\omega''$. The frequency $\omega^{i}_{2}$ defines the end of the absorption band, whose width is shortened by the presence of magnetic conductivity. At the same time, the real part of the indices becomes finite (positive or negative), allowing the signal propagation, $V_{E}^{\pm}\neq 0$ (forbidden in this interval for a usual Drude dielectric).

\item For $\omega^{i}_{2} <\omega < \omega''$, there is no absorption as one can see in the dotted curves of Figs. \ref{plot-n-mais-isotropic-case-Drude-with-magnetic-current} and \ref{plot-n-menos-isotropic-case-Drude-with-magnetic-current}. Consequently, the energy velocities $V_{E}^{\pm}$ start to increase progressively to their asymptotic values (which coincide with the usual case ones). Furthermore, in this range one still has $V_{E}^{-} <0$, which is in agreement with the negative refraction of $n_{-}$ that occurs for $\omega^{i}_{1} <\omega < \omega''$.

\item For $\omega >\omega^{i}_{2}$: it happens propagation without attenuation for both modes $n_{\pm}$.
\end{itemize}
One last comment, yet very subtle, concerns the similarity between the black solid curves in Figs.~\ref{group-velocity-isotropic-case} and \ref{plot-energy-velocity-isotropic-case-Drude-with-magnetic-current}. These curves represent the real part of the group velocity and the energy velocity, both for $\Sigma=0$, respectively.   Equations (\ref{energy-velocity-isotropic-7-0}) and (\ref{group-velocity-isotropic-case-7Re}) for $\Sigma=0$ read
\begin{subequations}
\begin{align}
V_{E}^{\pm}&= \frac{ \displaystyle  \sqrt{ \frac{| h(\omega)|}{2}} I_{+}}{\displaystyle \mu  C + \frac{\mu}{2} \sqrt{\epsilon'(\omega)^{2} + \epsilon''(\omega)^{2}} - \mu\epsilon'(\omega)/2} \; ,
\label{energy-velocity-isotropic-case-usual-scenario-1}
\\
\Re[v_{g}^{\pm}] &= \frac{ \displaystyle \sqrt{ \frac{ |h(\omega)|}{2} } \left( I_{+} C + I_{-} D\right)}{\displaystyle \mu \epsilon_{0} \left( C^{2} + D^{2} \right)} \; . \label{group-velocity-isotropic-case-usual-scenario-1}
\end{align}
\end{subequations}
These velocities are only equivalent when there is no absorption, $\gamma=0$. Hence, applying this condition, one finds $\epsilon''=0$ and $D=0$,
such that
\begin{align}
V_{E}^{\pm}-\Re[v_{g}^{\pm}] &= - \frac{\sqrt{ |\epsilon'(\omega) | + \epsilon'(\omega)}}{ C \sqrt{2 \mu \epsilon_{0} }}  \,  \frac{ |\epsilon'(\omega)| - \epsilon'(\omega)}{2C+ |\epsilon'(\omega)| - \epsilon'(\omega)} \; , \label{difference-between-group-and-energy-velocities-isotropic-usual-case-3}
\end{align}
where we have used
\begin{align}
I'_{+} &= \sqrt{1+ \mathrm{sgn}[\mu  \epsilon'(\omega) ]} \; . \label{difference-between-group-and-energy-velocities-isotropic-usual-case-2}
\end{align}
Obviously, it holds
\begin{align}
V_{E}^{\pm}-\Re[v_{g}^{\pm}] &= 0 \; , \label{difference-between-group-and-energy-velocities-isotropic-usual-case-4}
\end{align}
for $\epsilon'(\omega) >0$ or $\epsilon'(\omega) <0$. Note that one has: i) $\epsilon'(\omega) >0$ for $\omega<\omega_{0}$, or $\omega >\sqrt{\omega_{0}^{2} + \omega_{p}^{2}}$; ii) $\epsilon'(\omega) <0$, when $\omega_{0} < \omega < \sqrt{\omega_{0}^{2} + \omega_{p}^{2}}$.

\section{\label{section-antisymmetric-case}The case of antisymmetric magnetic conductivity tensor}
In this case, the antisymmetric conductivity tensor is parameterized in terms of a vector ${\bf{b}}$,
\begin{eqnarray}
\label{sigmaBsymm}
\sigma_{ij}^{B}=\epsilon_{ijk} \, b_{k} \; ,
\end{eqnarray}
implying an antisymmetric magnetic current density,
\begin{align}
\label{ASSYC}
{\bf{J}}={\bf{b}} \times {\bf{B}} \; ,
\end{align}
where $\bf{b}$ is a constant vector in space. Such an antisymmetric magnetic current was investigated in the context of Weyl semimetals \cite{Kaushik1}, where $\hat{\bf{b}}$ defines a direction in space. This kind of allowed transverse chiral magnetic photocurrent (orthogonal to the $\bf{B}$ field) is induced by linearly polarized light in WSMs in the presence of parity violation. The effective permittivity tensor is
\begin{eqnarray}
\label{barepsilonijantisymm}
\overline{\epsilon}_{ij}(\omega)= \left[ \, \epsilon(\omega)
- \frac{i}{\omega} ({\bf b}\cdot{\bf n}) \, \right]\delta_{ij}
+\frac{i}{\omega} \, n_{i} \, b_{j} \; ,
\end{eqnarray}
which replaced in \eqref{Fresnel-equation-1} yields the $n$-polynomial equation
\begin{eqnarray}\label{Eqsnantisymm}
{\bf n}^2+\frac{i\,\mu}{\omega} \, ({\bf b}\cdot{\bf n})-\mu\,\epsilon(\omega)
=0 \; ,
\end{eqnarray}
whose solution provides the dispersive refractive index
\begin{eqnarray}
\label{nantisymmetric}
n(\omega) = -\frac{i\,\mu}{2\omega} \, (\hat{{\bf k}}\cdot {\bf b})
+ \sqrt{\mu\epsilon(\omega)-\frac{\mu^2}{4\omega^2}(\hat{{\bf k}}\cdot {\bf b})^2 }
\; ,
\end{eqnarray}
where $\hat{\bf{k}}$ is the direction of propagation.

\begin{figure}[h]
\centering
\includegraphics[scale=0.68]{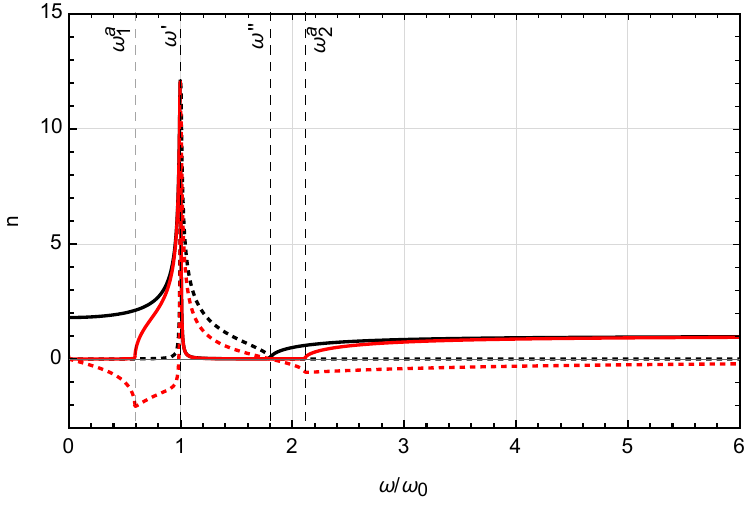}
\caption{Refractive index $n$ of \eqref{n-antisymmetric-simplified-1}. Solid (dotted) curves represent the real (imaginary) pieces of $n$. Here, we have used: $\mu=1$, $\epsilon_{0}=1$, $\omega_{p} /\omega_{0}=1.5$, $\gamma/\omega_{0}=0.01$, $b/\omega_{0}=2.5$, $\cos\theta=1$ (red). The black lines indicate the usual case.  The frequency  $\omega^{a}_{2}$ defines the end of the modified absorption zone, enlarged in relation to the one of a usual DD.}
\label{plot-n-antisymmetric-parallel-case-Drude-with-magnetic-current}
\end{figure}

\begin{figure}[h]
\centering
\includegraphics[scale=0.68]{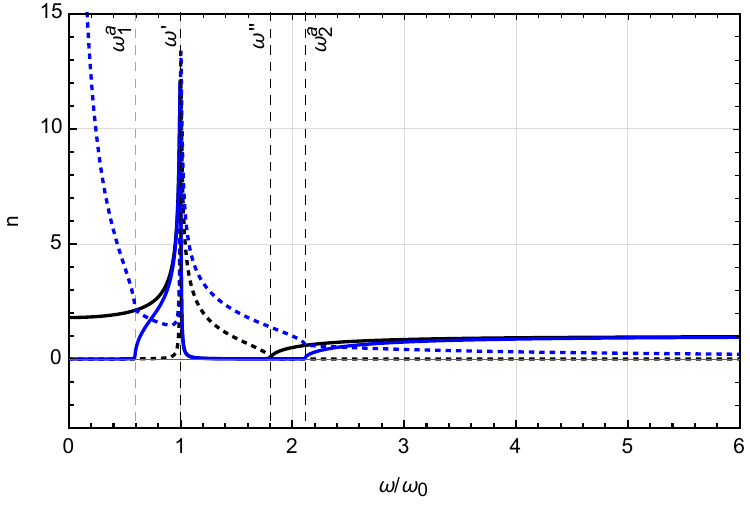}
\caption{Refractive index $n$ of \eqref{n-antisymmetric-simplified-1}. Solid (dotted) curves represent the real (imaginary) pieces of $n$. Here, we have used: $\mu=1$, $\epsilon_{0}=1$, $\omega_{p} /\omega_{0}=1.5$, $\gamma/\omega_{0}=0.01$, $b/\omega_{0}=2.5$, $\cos\theta=-1$ (blue). The black lines indicate the usual case. Here one also notes the enlargment of the absorption zone in relation to the one of a usual DD. }
\label{plot-n-antisymmetric-antiparallel-case-Drude-with-magnetic-current}
\end{figure}

Substituting the dielectric permittivity (\ref{epsilonomega}), the refractive index
(\ref{nantisymmetric}) can be rewritten as
\begin{subequations}
\begin{align}
\label{n-antisymmetric-simplified-1}
n(\omega)=\Re[n(\omega)]+i\,\Im[n(\omega)] \; ,
\end{align}
where the real and imaginary parts are
\begin{align}
\Re[n] &= \sqrt{ \frac{|f(\omega)|}{2} } \, A_{+} \; , \label{real-part-n-antysimmetric-1} \\
\Im[n] &= \sqrt{ \frac{|f(\omega)|}{2} } \, A_{-} - \frac{\mu}{2\omega} ( \hat{\bf{k}} \cdot {\bf{b}} ) \; , \label{imaginary-part-n-antisymmetric-1}
\end{align}
with $\epsilon'$ and $\epsilon''$ of Eqs.~(\ref{real-part-drude-epsilon-definition}) and (\ref{imaginary-part-drude-epsilon-definition}), respectively, and
\begin{align}
A_{\pm} &= \sqrt{\sqrt{  1 + \mu^2 \epsilon_{0}^2 \left[\frac{\epsilon''(\omega)}{f(\omega)}\right]^2 }  \pm \mbox{sgn}[f(\omega)] } \; , \label{A-funcions-definitions-1}
\end{align}
\end{subequations}
where $f(\omega)=\mu\epsilon^{\prime}(\omega)-\mu^2({\bf b}\cdot\hat{{\bf k}})^2/(2\omega)^2$.
The general behaviors of $\Re[n]$ and $\Im[n]$ are represented in Figs.~\ref{plot-n-antisymmetric-parallel-case-Drude-with-magnetic-current} and \ref{plot-n-antisymmetric-antiparallel-case-Drude-with-magnetic-current}, considering two distinct configurations of the vector $\bf{b}$ relative to the direction of propagation, $\cos\theta=\pm 1$, defining propagation parallel and anti-parallel to ${\bf{b}}$, respectively, with $\hat{\bf{k}}\cdot {\bf{b}} = b \cos\theta$. The case $\theta=\pi/2$ is effectively equivalent to the situation of a dielectric stripped of magnetic conductivity ($b=0$), which is represented by a black line in Figs.~\ref{plot-n-antisymmetric-parallel-case-Drude-with-magnetic-current} and \ref{plot-n-antisymmetric-antiparallel-case-Drude-with-magnetic-current}.

Near the origin, the real piece of $n$ goes to infinity for $\cos\theta=1$ whereas it goes to zero for the antiparallel propagation ($\cos\theta=-1$), a different behavior in comparison with the conventional DD, as already mentioned. We notice in Figs.~\ref{plot-n-antisymmetric-parallel-case-Drude-with-magnetic-current} and \ref{plot-n-antisymmetric-antiparallel-case-Drude-with-magnetic-current} that around $\omega/\omega_{0}=1$ the real parts of $n$ for parallel and anti-parallel propagation $(\cos\theta=\pm1)$ become equivalent and nearly overlap to the black line. This superposition also occurs in the very high-frequencies regime.

In the very high-frequency regime, the real part goes as $\Re[n](\omega \gg \omega_{0}) \rightarrow \sqrt{\mu\epsilon_{0}}$. For the very low-frequency limit, $\omega \rightarrow 0$, the refractive index becomes purely imaginary,
\begin{align}
n &\simeq \frac{ i \mu b}{2\omega} \left( -\cos\theta + |\cos \theta |\right) \; , \label{expansion-refractive-index-antisymmetric-case}
\end{align}
establishing $\Re[n]=0$ near the origin. Further, it yields $\Im[n] =0$ for $\cos\theta=1$ (see dashed line in Fig.~\ref{plot-n-antisymmetric-parallel-case-Drude-with-magnetic-current}) and a diverging $n$ for $\cos\theta=-1$ (see dashed line in Fig.~\ref{plot-n-antisymmetric-antiparallel-case-Drude-with-magnetic-current}).

Differently from the standard Drude-Lorentz model, where the absorption occurs around the resonance frequency $\omega_{0}$, the presence of a non-null magnetic conductivity, given by $b=|{\bf{b}}|$, enables absorption even in the small frequency regime, $\omega \ll \omega_{0}$, accordingly to \eqref{expansion-refractive-index-antisymmetric-case}. Therefore, in this antisymmetric scenario for the magnetic conductivity, absorption effects can occur in regions outside the neighborhood of the resonance frequency $\omega_{0}$. On the other hand, such an effect does not happen by means of absorption peak as it occurs in the DD.

The peaks of the real and imaginary parts are given by
\begin{subequations}
\label{n-antisymmetric-in-resonance-1}
\begin{align}
\Re[n](\omega=\omega_{0}) &= \sqrt{ \frac{ | f(\omega_{0})|}{2}} A_{+}(\omega_{0}) \; ,     \label{n-antisymmetric-in-resonance-2}\\
\Im[n](\omega=\omega_{0}) &= \sqrt{ \frac{ | f(\omega_{0})|}{2}} A_{-}(\omega_{0}) - \frac{\mu}{2\omega_{0}}(\hat{\bf{k}}\cdot {\bf{b}}) \; ,  \label{n-antisymmetric-in-resonance-3}
\end{align}
\end{subequations}
where $f(\omega_0)=\mu\epsilon_0-\mu^2({\bf b}\cdot\hat{\bf k})^2/(2\omega_0)^2$. A very subtle difference between the absorption peaks (imaginary part of $n$) can be observed by comparing Figs.~\ref{plot-n-antisymmetric-parallel-case-Drude-with-magnetic-current} and \ref{plot-n-antisymmetric-antiparallel-case-Drude-with-magnetic-current}. One notices that $\Im[n]|_{\theta=0} < \Im[n]|_{\theta=\pi}$, a direct consequence of the last term in \eqref{n-antisymmetric-in-resonance-3}, which reveals an observable distinction regarding the propagation direction relative to the magnetic conductivity vector ${\bf{b}}$.
%

\subsection{Propagation modes}

%
%
%
The propagation modes are found for ${\bf{n}}$ along the ${\cal Z}$-axis, and the vector $\mathbf{b}$ written as
\begin{equation}
\mathbf{n}=(0,0,n)\,,\quad \mathbf{b}=b\left(0,\sin\theta,\cos\theta\right)\equiv (0,b_2,b_3) \; ,
\label{COORD}
\end{equation}
where $\theta$ is the angle between $\mathbf{n}$ and $\mathbf{b}$. The matrix $M_{ij}$ of \eqref{Fresnel-equation-2} is
\begin{equation}
[ M_{ij}]=\left[
\begin{array}{ccc}
n^{2}-\mu\epsilon + n \frac{ i\mu  }{\omega}b_3 & 0 & 0
\\
\\
0 & n^{2}-\mu\epsilon + n \frac{i\mu }{\omega}b_3 & 0
\\
\\
0 & -n \frac{i \mu  }{\omega}b_2 & -\mu\epsilon \\
\end{array}
\right] \; ,
\label{MATRIXb}
\end{equation}
with $\mathbf{n\cdot b}=n\, b_3$. Replacing the index relation (\ref{Eqsnantisymm}) in the matrix (\ref{MATRIXb}), the condition $M_{ij}E_j=0$ provides two mutually orthogonal propagation modes
\begin{equation}
\mathbf{E}_{\pm}=\frac{1}{\sqrt{2(1+|Q|^{2})}}\begin{pmatrix}
\pm\sqrt{1+|Q|^{2}} \\
-1 \\
\mathrm{i} Q e^{i\alpha}\\
\end{pmatrix} \; ,
\label{modesantisymmetric}
\end{equation}
with $Q=b_{2}N/ (\omega \epsilon)$, endowed however with a longitudinal component. Here, we have rewritten the complex refractive index as $n=Ne^{i\alpha}$, with
\begin{align}
N&=\sqrt{ \left( \sqrt{ \frac{ |f(\omega)|}{2}}A_{+} \right)^{2} + \left( \sqrt{ \frac{ |f(\omega)|}{2}}A_{-} - \frac{\mu b\cos\theta}{2\omega} \right)^{2}} \; ,
\end{align}
and
\begin{align}
\tan \alpha &=\frac{A_{-}}{A_{+}}- \frac{\mu b \cos\theta}{\omega A_{+} \sqrt{2 |f(\omega)|}} \; .
\end{align}
For the particular case where $b_{2}=0$, the solutions in \eqref{modesantisymmetric} simplify as
\begin{align}
{\mathbf{E}}_{\pm}=\frac{1}{\sqrt{2}}\begin{pmatrix}
\pm 1 \\
-1 \\
0
\end{pmatrix} \; , \label{modesantisymmetric-simplified}
\end{align}
representing linearly polarized vectors, orthogonal to the propagation direction. As the refractive index of \eqref{n-antisymmetric-simplified-1} has an imaginary piece, there is absorption for both modes given in \eqref{modesantisymmetric-simplified} (in equal magnitude), measured in terms of the absorption coefficient \cite{Zangwill}, $\tilde{\alpha}=2\omega \, \mathrm{Im}(n) =\mu \, b_{3}$.

\subsection{The group and energy velocities of the antisymmetric case}

Rewritting Eq. (\ref{Eqsnantisymm}) in terms of ${\bf k}$, one has
\begin{eqnarray}\label{eqkantisymm}
{\bf k}^2+i\,\mu \left({\bf b}\cdot{\bf k}\right)-\mu\epsilon(\omega)\,\omega^2 = 0 \; ,
\end{eqnarray}
whose solution yields the dispersion relation
\begin{eqnarray}
k(\omega) = -\frac{i\mu}{2}( {\bf b}\cdot\hat{{\bf k}})+\sqrt{\mu\epsilon(\omega)\,\omega^2
-\frac{\mu^2}{4} ( {\bf b}\cdot\hat{{\bf k}} )^2 } \; ,\label{solkantisim}
\end{eqnarray}
with real and imaginary parts given as, respectively,
\begin{subequations}
\label{real-imaginary-part-of-momentum-antisymmetric-case-1}
\begin{align}
k^{\prime}(\omega) &= \omega \, \sqrt{ \frac{|f(\omega)|}{2} } \, A_{+} \; ,
\label{Realkpantisymm} \\
k^{\prime\prime}(\omega) &= \omega \, \sqrt{ \frac{|f(\omega)|}{2} } \, A_{-} - \frac{\mu}{2} ( \hat{\bf{k}} \cdot {\bf{b}} ) \; .
\label{Imkpantisymm}
\end{align}
\end{subequations}
Taking the derivative in relation to $k^{i}$-component in (\ref{eqkantisymm}), the $i$th-component of the group velocity is
\begin{align}
\label{vgantisim}
v_{g}^{i}=\frac{ k^{i}+i\,\mu\,b^{i}/2}{\mu\omega\left[ \epsilon(\omega)+f_{\epsilon}(\omega) \, \omega/2 \right]} \; .
\end{align}
Using now $k^{i}=k^{i\prime} + i k^{i \prime\prime}$, one finds
\begin{align}
	\label{vgantisimetric}
	{\bf v}_{g} &= \frac{ {\bf k}^{\prime}/(\mu\omega) + i \, \omega \epsilon'' \hat{{\bf k}}'/{(2 k')}} { C + i D }  \; ,
\end{align}
with $C$ and $D$ defined by \eqref{C-D-definitions}. The latter also indicates that the group velocity is defined along the ${\bf{k}}'$ direction. Substituting the solution
(\ref{real-imaginary-part-of-momentum-antisymmetric-case-1}), the group velocity is written as a function of $\omega$ :

\begin{subequations}
\label{real-imaginary-parts-group-velocity-antisymmetric-case-1}
\begin{align}
\Re[v_{g}] &= \frac{ (\mu\epsilon_{0})^{-1}}{C^{2} + D^{2}} \sqrt{ \frac{ |f(\omega)|}{2}} \left( A_{+} C + A_{-} D \right) \; , \label{real-imaginary-parts-group-velocity-antisymmetric-case-2} \\
\Im[v_{g}] &= \frac{ (\mu\epsilon_{0})^{-1}}{C^{2}+D^{2}} \sqrt{ \frac{ |f(\omega)|}{2}}  \left( A_{-} C - A_{+} D \right) \; . \label{real-imaginary-parts-group-velocity-antisymmetric-case-3}
\end{align}
\end{subequations}

\begin{figure}[h]
\begin{centering}
\includegraphics[scale=0.68]{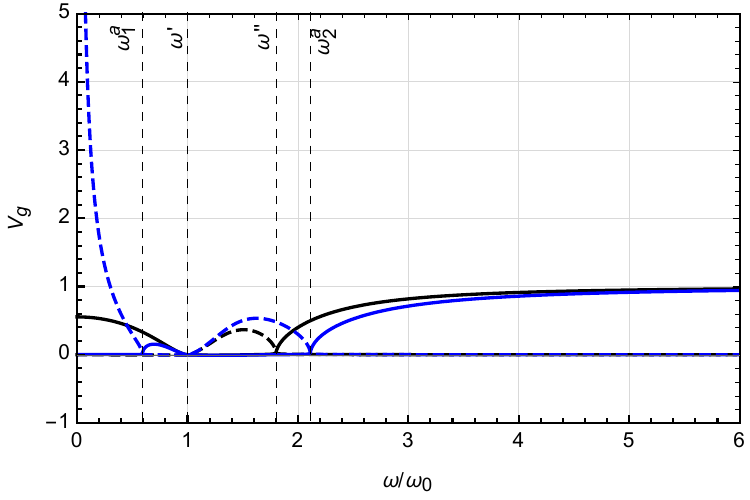}
\par\end{centering}
\caption{\label{plot-group-velocity-antisymmetric-case} Group velocity of \eqref{real-imaginary-parts-group-velocity-antisymmetric-case-1} for $\cos\theta=\pm 1$. The solid (dashed) lines represent the real (imaginary) parts of $v_{g}$. Here, we have used: $\mu=1$, $\epsilon_{0}=1$, $\omega_{p} /\omega_{0}=1.5$, $\gamma/\omega_{0}=0.01$, $b/\omega_{0}=2.5$ (blue), $b=0$ (black, the usual case). The vertical dashed lines indicate $\omega / \omega_{0} \approx \{0.59, 1, 1.8, 2.11\}$ from left to right, accordingly with \eqref{frequencies-antisymmetric-case}.}
\end{figure}

The group velocity behavior as function of the dimensionless parameter $\omega/\omega_{0}$ is shown in Fig.~\ref{plot-group-velocity-antisymmetric-case}. For high-frequencies, it holds $\Re[v_{g}]= 1/\sqrt{\mu_0\epsilon_0}=1$ (for $\mu=\mu_0$),
whereas the imaginary part goes to zero, $\Im[v_{g}]=0$. At the origin, its real piece is well behaved whereas the imaginary part goes to infinity. This latter behavior is also observed in a dielectric with constant magnetic conductivity, $\epsilon^{\prime}=cte$, $\epsilon^{\prime\prime}=\sigma/\omega$, $\Sigma=cte$, as seen in Ref. \cite{PedroMario}.

Assessing now the energy propagation signal, we implement (\ref{Realkpantisymm}) in (\ref{VEresult}), in such a way the energy velocity is
\begin{subequations}
\begin{align}
V_{E}&=  \sqrt{ \frac{ |f(\omega)|}{2}} \frac{A_{+}} {U^{A}} \; , \label{energy-velocity-antisymmetric-5}
\end{align}
with
\begin{align}
U^{A}&= \frac{\mu\epsilon'}{2} + \frac{\mu \omega}{2} \frac{ \partial \epsilon'}{\partial \omega} + \frac{ |f(\omega)|}{4} (A_{+}^{2}+A_{-}^{2}) + \frac{ \mu^{2}b^{2}}{8\omega^{2}} \cos^{2}\theta+
\nonumber \\
&\phantom{=} - \frac{\mu b\cos\theta}{2\omega} A_{-} \sqrt{ \frac{ |f(\omega)|}{2}} \; . \label{energy-velocity-antisymmetric-5-1}
\end{align}
\end{subequations}

\begin{figure}[h]
\begin{centering}
\includegraphics[scale=0.68]{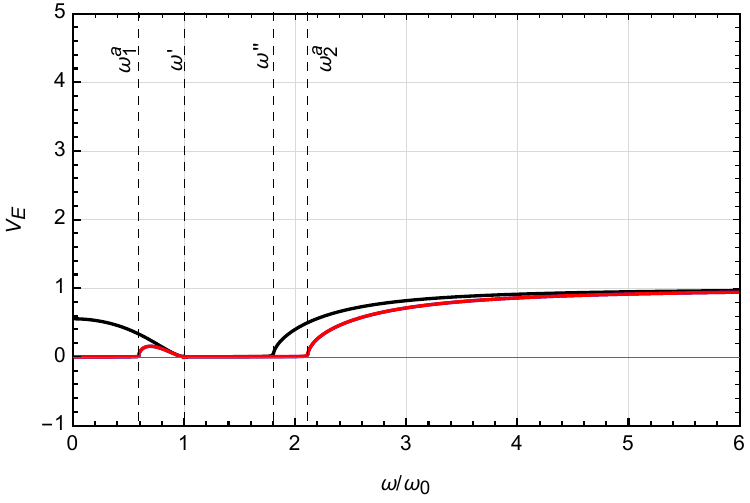}
\par\end{centering}
\caption{\label{plot-energy-velocity-antisymmetric-case-Drude-with-magnetic-current} Energy velocity of \eqref{energy-velocity-antisymmetric-5} in terms of the dimensionless parameter $\omega/\omega_{0}$. The solid red line represents $V^{E}$ for $\cos\theta=\pm 1$. The black solid curve 
	illustrates the usual case (\ref{VEresult}), where $b=0$. Here, we have used: $\mu=1$, $\epsilon_{0}=1$, $\omega_{p} /\omega_{0}=1.5$, $\gamma/\omega_{0}=0.01$, and $b/\omega_{0}=2.5$. The vertical dashed lines indicate $\omega / \omega_{0} \approx \{0.59, 1, 1.8, 2.11\}$, from left to right, accordingly with \eqref{frequencies-antisymmetric-case}.}
\end{figure}

The behavior of the energy velocity is depicted in Fig.~\ref{plot-energy-velocity-antisymmetric-case-Drude-with-magnetic-current} in terms of $\omega/\omega_{0}$. The frequencies $\omega^{a}_{1, 2}$ define a region where occurs an absorption peak followed by a strong reduction of the real part of the refractive index $n$, as indicated in Figs.~\ref{plot-n-antisymmetric-parallel-case-Drude-with-magnetic-current} and \ref{plot-n-antisymmetric-antiparallel-case-Drude-with-magnetic-current}. In details, we observe that:

\begin{itemize}

\item For $0<\omega < \omega^{a}_{1}$: one has a total absorption range, since $\mathrm{Re}[n]=0$, where there is no propagation. Hence, the energy velocity is null in this frequency window.

\item For $\omega^{a}_{1} <\omega < \omega'$: there occurs propagation together with absorption, since $\mathrm{Re}[n] \neq 0$ and $\mathrm{Im}[n] \neq 0$. As a consequence, there is a non-null energy velocity in this small region, as indicated in Fig.~\ref{plot-energy-velocity-antisymmetric-case-Drude-with-magnetic-current}. As the frequency approaches $\omega'$, there occurs an absorption peak followed by a rapid decreasing of $\mathrm{Re}[n]$ to zero, which marks the beginning of a second total absorption zone, where $V_{E}$ vanishes. This behavior appears in terms of a diminishing $V_{E}$ for $\omega$ approaching $\omega'$ and $V_{E}=0$ thereafter. See Fig.~\ref{plot-energy-velocity-antisymmetric-case-Drude-with-magnetic-current}.

\item For $\omega'<\omega<\omega^{a}_{2}$: there is no propagation, $\mathrm{Re}[n]=0$, as pointed in Figs.~\ref{plot-n-antisymmetric-parallel-case-Drude-with-magnetic-current} and \ref{plot-n-antisymmetric-antiparallel-case-Drude-with-magnetic-current}, in such a way only absorption takes place and $V_{E}=0$. This interval magnitude represents the enlargement of the absorption zone in the presence of the antisymmetric magnetic conductivity, compared to the usual DD absorption window, which is now incremented by $\Delta \omega=\omega^{a}_{2} - \omega''$.

\item For $\omega> \omega^{a}_{2}$: the absorption starts to diminish while the propagation regime is partially restored. As a consequence, the energy velocity assumes increasing magnitude, tending to its asymptotic value.

\item The entire plot of Fig.~\ref{plot-energy-velocity-antisymmetric-case-Drude-with-magnetic-current} reveals a scenario where the signal propagation is severely constrained for frequencies below $\omega^{a}_{2}$, which is in contrast with the signal propagation in the absorption windows of the Drude isotropic conductivity case, represented by the energy velocity of Fig.~\ref{plot-energy-velocity-isotropic-case-Drude-with-magnetic-current}. 

\end{itemize}

The characteristic frequencies can be determined by solving the sixth-order equation in $\omega$,
\begin{eqnarray}
&&
\gamma^{2} \left( \omega^{4} - \frac{\mu b^{2}\cos^{2}\theta}{4\epsilon_{0}}  \omega_{0}^{2} \right) +(\omega_{0}^{2}-\omega^{2}) \times
\nonumber \\
&&
\times  \left[ \omega^{2} \left(\omega_{0}^{2}-\omega^{2}+\omega_{p}^{2} \right) - \frac{\mu b^{2}\cos^{2}\theta}{4\epsilon_{0}} \left(\omega_{0}^{2}-\omega^{2}-\gamma^{2} \right) \right] = 0 \; ,
\nonumber \\
\label{frequencies-antisymmetric-case}
\end{eqnarray}
from which the real and positive solutions yield $\omega^{a}_{1, 2, 3}$ when $b\cos\theta \neq 0$, and $\omega^{\prime, \prime\prime}$ when $b\cos\theta = 0$. However, as $\omega^{a}_{3} \approx \omega'$, we have chosen to depict only $\omega'$ in the plots.

\section{\label{section-symmetric-case}The case of the symmetric conductivity tensor}
\begin{figure}[H]
\centering
\includegraphics[scale=0.68]{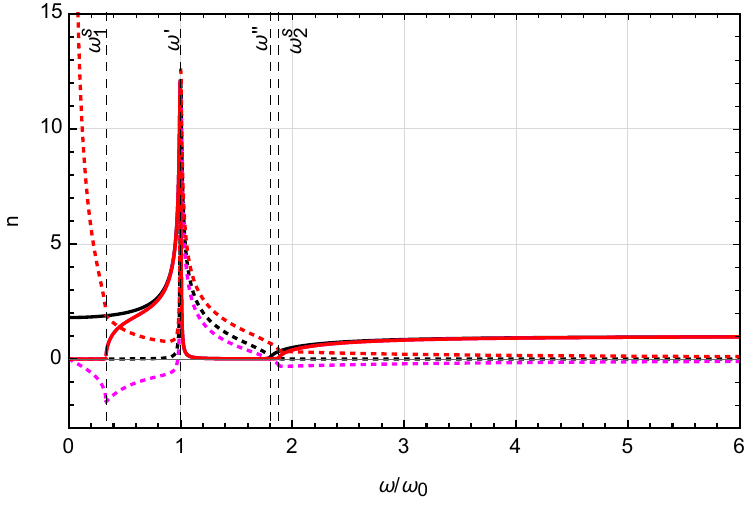}
\caption{
Refractive indices $n_{\pm}$ of \eqref{n-symmetric-case-0}. Solid (dotted) lines indicate the real (imaginary) part of $n_{\pm}$. The magenta (red) curves illustrate $n_{\pm}$ for $\cos\varphi=1$, respectively. The black lines depicts the usual case where $|{\bf{a}}\times {\bf{c}}|=0$. Here, we have used: $\mu=1$, $\epsilon_{0}=1$, $\omega_{p} /\omega_{0}=1.5$, $\gamma/\omega_{0}=0.01$, and $ |{\bf{a}}\times {\bf{c}}|/\omega_{0}=2.5$. The solid red and magenta curves lie on top of each other. The plot also reveals the opening of a new absorption band for low frequencies, while the dominant lossy region width is slightly augmented. }  
\label{plot-n-parallel-symmetric-case-Drude-with-magnetic-current}
\end{figure}

\begin{figure}[h]
\centering
\includegraphics[scale=0.68]{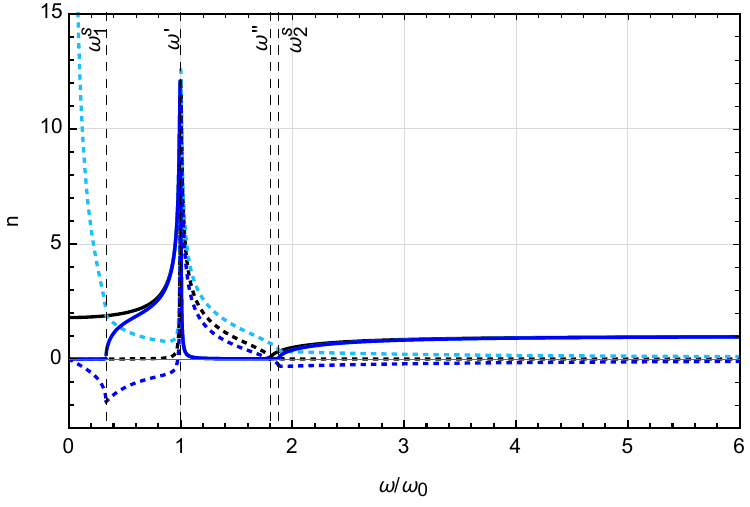}
\caption{Refractive indices $n_{\pm}$ of \eqref{n-symmetric-case-0}. Solid (dotted) lines indicate the real (imaginary) part of $n_{\pm}$. The cyan (blue) lines represent $n_{\pm}$ for $\cos\varphi=-1$, respectively. The black lines illustrates the usual case, where $|{\bf{a}}\times {\bf{c}}|=0$. Here, we have used: $\mu=1$, $\epsilon_{0}=1$, $\omega_{p} /\omega_{0}=1.5$, $\gamma/\omega_{0}=0.01$, and $ |{\bf{a}}\times {\bf{c}}|/\omega_{0}=2.5$. The solid blue and cyan curves are overlapped. One notes the opening of a new absorption band for low frequencies, while the dominant lossy window length is slightly expanded. } 
\label{plot-n-antiparallel-symmetric-case-Drude-with-magnetic-current}
\end{figure}

We write the magnetic conductivity tensor in the symmetric form
\begin{eqnarray}\label{sigmaBsymm}
\sigma_{Bij}=\frac{1}{2} \left( a_{i} \, c_{j} + a_{j} \, c_{i} \right) \; ,
\end{eqnarray}
where $a_{i}$ and $c_{j}$ are components of the orthogonal background vectors ${\bf a}$ and ${\bf c}$, respectively.
In this case, the permittivity tensor (\ref{barepsilonij}) is
\begin{eqnarray}\label{barepsilonijsymm}
\overline{\epsilon}_{ij}(\omega)= \epsilon(\omega) \delta_{ij}
+\frac{i}{2\omega} \left[ a_{i} \left( {\bf c} \times {\bf n} \right)_{j}
+ c_{i} \left( {\bf a} \times {\bf n} \right)_{j}\right]. \; \;
\end{eqnarray}
The null determinant of the matrix $M_{ij}$ with permittivity (\ref{barepsilonijsymm}) provides $n$-polynomial equations,
\begin{align}
n^2 \, - \, \mu \,\epsilon(\omega) \pm \frac{i\,\mu}{2\omega} \, {\bf n}\cdot({\bf a}\times{\bf c})
&= 0 \; .
\label{Eqsnsymm1}
\end{align}
Writing ${\bf n}=n \, \hat{{\bf k}}$, the solutions of (\ref{Eqsnsymm1}) are
\begin{align}
n_{\pm}(\omega) &=\mp\frac{i\,\mu}{4\omega} \, \hat{{\bf k}}\cdot({\bf a}\times{\bf c})
+ \sqrt{\mu\epsilon(\omega)-\frac{\mu^2}{16\omega^2}[\hat{{\bf k}}\cdot({\bf a}\times{\bf c})]^2 } \; ,
\label{n1sym}
\end{align}
or
\begin{equation}\label{n-symmetric-case-0}
n_{\pm}= \Re[n_{\pm}]+ i \, \Im[n_{\pm}] \; ,
\end{equation}
with
\begin{subequations}
\begin{align}
\Re[n_{\pm}] &= \sqrt{ \frac{|g(\omega)|}{2}  } S_{+}  \; , \label{real-part-n-simmetric-1} \\
\Im[n_{\pm}] &= \sqrt{ \frac{|g(\omega)|}{2} } \, S_{-} \mp \frac{\mu}{4\omega} \, \hat{\bf{k}} \cdot ( {\bf{a}} \times {\bf c} ) \; , \label{imaginary-part-n-symmetric-1}
\end{align}
\end{subequations}
where
\begin{align}
S_{\pm} &= \sqrt{\sqrt{  1 + \mu^2  \left[\frac{ \epsilon^{\prime\prime}(\omega) }{ g(\omega) } \right]^2 } \pm \mbox{sgn}[g(\omega)] } \; , \label{B-functions-definitions-1}
\end{align}
and 
	\begin{align}
	g(\omega)= \mu\epsilon^{\prime}(\omega) - \mu^{2}(\hat{\bf{k}}\cdot ({\bf{a}}\times{\bf c}) )^{2}/(4\omega)^2.
\end{align}
Notice that the expressions (\ref{real-part-n-simmetric-1}) and (\ref{imaginary-part-n-symmetric-1}) can be obtained from the results
of the antisymmetric case (\ref{real-part-n-antysimmetric-1}) and (\ref{imaginary-part-n-antisymmetric-1}) by implementing ${\bf b} \rightarrow \pm ({\bf a}\times{\bf c})/2$. The behavior of $n_{\pm}$ as function of $\omega/\omega_{0}$ is illustrated in Figs.~\ref{plot-n-parallel-symmetric-case-Drude-with-magnetic-current} and \ref{plot-n-antiparallel-symmetric-case-Drude-with-magnetic-current}, where we have used $\hat{\bf{k}} \cdot ({\bf{a}}\times {\bf{c}})= |{\bf{a}}\times {\bf{c}}| \cos\varphi$. From \eqref{n-symmetric-case-0}, one notices that $\Re[n_{+}]=\Re[n_{-}]$ and $\Im[n_{\pm}]|_{\varphi=0, \pi} = \Im[n_{\mp}]|_{\varphi=\pi, 0}$, which indicates a kind of correspondence symmetry between the absorptive terms of $n_{\pm}$ relative to the sense (positive or negative) of propagation.

%
In the very high-frequency limit, the refractive indices become $n_{\pm} \rightarrow \sqrt{\mu\epsilon_{0}}$. On the other hand, in the low-frequency regime, one obtains
\begin{align}
n_{\pm} \simeq  \frac{i\mu |{\bf{a}}\times {\bf{c}}|}{4\omega} \left(\mp \cos\varphi + |\cos\varphi | \right) \; .
\label{n-symmetric-expansion-0}
\end{align}
Hence $n_{+}|_{\varphi=0}=n_{-}|_{\varphi=\pi}=0$ and $n_{+}|_{\varphi=\pi}=n_{-}|_{\varphi=0} \simeq i \mu |{\bf{a}}\times {\bf{c}}| / (2\omega)$, which explains the divergence in the origin observed in Figs.~\ref{plot-n-parallel-symmetric-case-Drude-with-magnetic-current} and \ref{plot-n-antiparallel-symmetric-case-Drude-with-magnetic-current}.


\subsection{Propagation modes}

To evaluate the propagating modes, we choose an appropriate coordinate system where the magnetic conductivity vectors are given by
${\bf{a}}=(a_{1}, 0, a_{3} )$, ${\bf{c}} = (0, c_{2}, 0)$, such that
\begin{align}
{\bf{a}}\times {\bf{c}}= (-a_{3}c_{2}, 0, a_{1}c_{2} ) \; .
\label{symmetric-propagating-3}
\end{align}

For propagation along the ${\cal Z}$-axis, ${\bf{n}}=(0, 0, n)$, the matrix $M_{ij}$ is
\begin{align}
[M_{ij}] &= \left[
\begin{array}{ccc}
n^{2}-\mu{\epsilon} - i \frac{\mu}{2\omega} a_{1}c_{2} n & 0 & 0
\\
\\
0   &    n^{2}-\mu{\epsilon} + i \frac{\mu}{2\omega} a_{1}c_{2} & 0
\\
\\
- i \frac{\mu}{2\omega} a_{3}c_{2} n  & 0 &  - \mu{\epsilon}
\end{array}
\right], \label{symmetric-propagating-4}
\end{align}
where ${\epsilon=\epsilon(\omega)}$ is the complex electric permittivity. Thus, by implementing \eqref{Eqsnsymm1} in \eqref{symmetric-propagating-4}, the relation $M_{ij}E_{j}=0$ provides the following polarization vectors:
\begin{align}
{\bf{E}}_{+} = \frac{1}{\sqrt{1+|A|^2}}  \begin{pmatrix}
1 \\
0 \\
-i A
\end{pmatrix} \, ,
\quad
{\bf{E}}_{-} = \begin{pmatrix}
	0\\
	1\\
	0
\end{pmatrix},
\label{symmetric-propagating-11}
\end{align}
associated with $n_{\pm}$, respectivelly, with $A={a_{3}c_{2}}/{(2\omega{\epsilon}n )}$. We observe that ${\bf{E}}_{\pm}$ represent
linearly polarized vectors, with ${\bf{E}}_{+}$ endowed with a longitudinal component.
%
Since the the real peace of $n_{\pm}$ is the same one, there is no birefringence. On the other hand, the imaginary pieces are different, meaning that the propagating modes are absorbed at different degrees, which can be measured by the absorption difference per unit length,
\begin{align}
\frac{\Delta_{d}}{l}=\frac{2\pi}{\lambda_0} \, \left[ \, \mathrm{Im}(n_{+})-\mathrm{Im}(n_{-}) \, \right] \; ,
\label{phase-shift1}
\end{align}
with $\lambda_{0}$ being the vacuum wavelength of incident light. Hence, for the indices of \eqref{n-symmetric-case-0} and propagation along the $z$-axis, the absorption difference is
\begin{align}
\frac{\Delta_{d}}{l}=-\frac{\mu a_{1} \, c_{2}}{2} \; .
\label{dicro}
\end{align}

%
\subsection{The group and energy velocities of the symmetric case}
Backing to \eqref{Eqsnsymm1} in terms of ${\bf k}$ and $\omega$, we rewrite it as
%
\begin{eqnarray}
k^2 \, \pm \, \frac{i\,\mu}{2} \, {\bf k}\cdot({\bf a}\times{\bf c}) \, - \, \mu \, \omega^2\,\epsilon(\omega)
&= 0 \; ,
\label{Eqsnsymm1k}
\end{eqnarray}
whose solutions are
\begin{equation}
k_{\pm}(\omega) = k'_{\pm} + i k''_{\pm} \; ,
\label{solksim1}
\end{equation}
with
\begin{subequations}
\label{real-and-imaginary-parts-momentum-symmetric-case-1}
\begin{eqnarray}
k^{\prime}_{\pm}(\omega) &=& \omega\,\sqrt{ \frac{|g(\omega)|}{2}  } \, S_{+} \; ,
\label{Reksymm} \\
k^{\prime\prime}_{\pm}(\omega) &=& \omega\,\sqrt{ \frac{|g(\omega)|}{2} } \, S_{-} \mp \frac{\mu}{4} \, \hat{\bf{k}} \cdot ( {\bf{a}} \times {\bf c} ) \; , \label{Imksymm}
\end{eqnarray}
\end{subequations}
and $S_{\pm}$ of \eqref{B-functions-definitions-1}. Evaluating the derivative of \eqref{Eqsnsymm1k} relative to $k^{i}$-component, one finds the following group velocities:
\begin{align}
\label{symmetric-case-14}
{\bf v}_{g}^{\pm} = \frac{\partial \omega}{\partial {\bf k}} = \frac{ \displaystyle \frac{{\bf k}}{\mu \omega} \pm i \, \frac{({\bf a \times c})}{4\omega}}{ \displaystyle \left(\epsilon' + \frac{\omega}{2} \frac{ \partial \epsilon'}{\partial \omega} \right) + i \left(\epsilon'' + \frac{\omega}{2} \frac{ \partial \epsilon''}{\partial \omega} \right)} \; .
\end{align}
Using \eqref{solksim1} the group velocity (\ref{symmetric-case-14}) takes on the form,
\begin{align}
\label{symmetric-case-15}
{\bf v}_{g}^{\pm} &= \frac{ \displaystyle \frac{{\bf k}^{\prime}}{\mu\omega} + i \, \frac{\omega \epsilon'' \, \hat{{\bf k}} }{2 k'}} { \displaystyle \left( \epsilon' + \frac{\omega}{2} \frac{\partial \epsilon'}{\partial \omega} \right) + i \left(\epsilon'' + \frac{\omega}{2} \frac{\partial \epsilon''}{\partial \omega} \right) }  \; .
\end{align}
whose real and imaginary pieces are
\begin{subequations}
\label{real-and-imaginary-parts-group-velocity-symmetric-case-1}
\begin{align}
\Re[ v^{\pm}_{g}] &= \frac{ (\mu\epsilon_{0})^{-1}}{C^{2} + D^{2}} \sqrt{ \frac{ |g(\omega)|}{2}} \left( S_{+} C + S_{-} D \right) \; ,  \label{real-and-imaginary-parts-group-velocity-symmetric-case-2} \\
\Im[v^{\pm}_{g}] &= \frac{ (\mu\epsilon_{0})^{-1}}{C^{2} + D^{2}} \sqrt{ \frac{ |g(\omega)|}{2}}  \left( S_{-} C - S_{+} D\right) \; .   \label{real-and-imaginary-parts-group-velocity-symmetric-case-3}
\end{align}
\end{subequations}

\begin{figure}[H]
\begin{centering}
\includegraphics[scale=0.68]{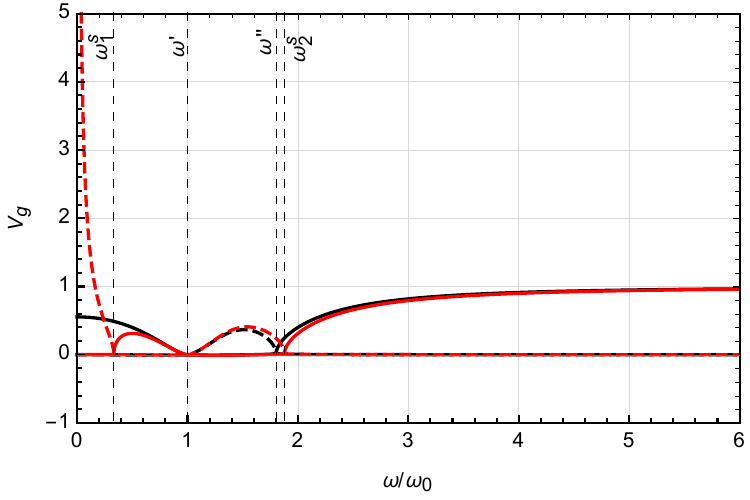}
\par\end{centering}
\caption{\label{group-velocity-symmetric-case} Group velocity of \eqref{real-and-imaginary-parts-group-velocity-symmetric-case-1} in terms of the dimensionless parameter $\omega/\omega_{0}$ for $\cos\varphi=\pm 1$. Solid (dashed) lines represent the real (imaginary) parts of $v_{g}$. The solid blue line represents $V_{E}^{\pm}$ for $\cos\theta=\pm 1$. Here, we have used: $\mu=1$, $\epsilon_{0}=1$, $\omega_{p} /\omega_{0}=1.5$, $\gamma/\omega_{0}=0.01$, $|{\bf{a}}\times {\bf{c}}|/\omega_{0}=2.5$ (red), and $|{\bf{a}}\times {\bf{c}}|/\omega_{0}=0$ (black). The vertical dashed lines indicate $\omega / \omega_{0} \approx \{0.33, 1, 1.8, 1.87\}$, from left to right, accordingly with \eqref{frequencies-symmetric-case}. }
\end{figure}

Figure \ref{group-velocity-symmetric-case} shows the real and imaginary parts of the group velocity given in \eqref{real-and-imaginary-parts-group-velocity-symmetric-case-1} in terms of the dimensionless parameter $\omega/\omega_{0}$. We have considered the previous case where ${\bf a}=(a_1,0,a_3)$ and ${\bf c}=(0,c_2,0)$, such that ${\bf a}\times{\bf c}=(-a_3c_2,0,a_1c_2)$, and $\hat{{\bf k}}=\hat{{\bf z}}$. For high frequencies $(\omega \gg \omega_0)$, one has $\Re[v_{g}] \rightarrow  1/\sqrt{\mu_0\epsilon_0}= 1$ and $\Im[v_{g}]=0$.

Considering $a_{3}=0$, the propagating modes represented by \eqref{symmetric-propagating-11} turn out perpendicular to the direction of propagation, satisfying the conditions (\ref{complex-group-velocity-isotropic-54A}) and (\ref{complex-group-velocity-isotropic-54B}) for the evaluation of the energy velocity. In this sense, Eqs. (\ref{Reksymm}) and (\ref{Imksymm}) yield
\begin{subequations}
	\begin{align}
		V_{E}^{\pm} &= \sqrt{ \frac{ |g(\omega)|}{2}} \frac{S_{+}}{U^{S}_{\pm}} \; , \label{energy-velocity-symmetric-4}
	\end{align}
	with
	\begin{align}
		U^{S}_{\pm}&= \frac{\mu \epsilon'}{2}+\frac{\mu\omega}{2} \frac{ \partial \epsilon'}{\partial \omega} + \frac{ |g(\omega)|}{4} (S_{+}^{2}+S_{-}^{2}) + \nonumber \\
		&\phantom{=}+ \frac{ \mu^{2}|{\bf{a}}\times {\bf{c}}|^{2}}{32\omega^{2}} \cos^{2}\varphi \pm \frac{|{\bf{a}}\times {\bf{c}}|}{4\omega} \cos\varphi \sqrt{ \frac{ |g(\omega)|}{2}} S_{-} \; . \label{energy-velocity-symmetric-4-1}
	\end{align}
\end{subequations}

\begin{figure}[H]
\begin{centering}
\includegraphics[scale=0.68]{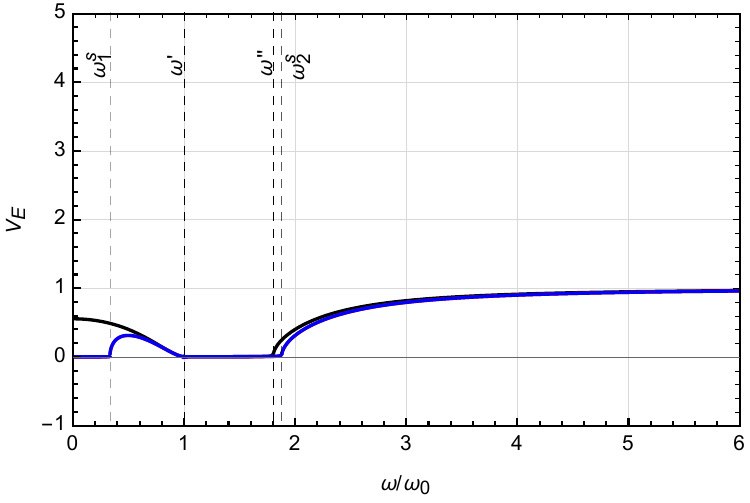}
\par\end{centering}
\caption{\label{plot-energy-velocity-symmetric-case-Drude-with-magnetic-current} Energy velocity $V_{E}^{\pm}$ of \eqref{energy-velocity-symmetric-4} in terms of the dimensionless parameter $\omega/\omega_{0}$. The solid blue line represents $V_{E}^{\pm}$ for $\cos\theta=\pm 1$. The black solid curve depicts the usual case (null magnetic conductivity) of \eqref{VEresult}. Here, we have used: $\mu=1$, $\epsilon_{0}=1$, $\omega_{p} /\omega_{0}=1.5$, $\gamma/\omega_{0}=0.01$, and $|{\bf{a}}\times {\bf{c}}|/\omega_{0}=2.5$. The vertical dashed lines indicate $\omega / \omega_{0} \approx \{0.33, 1, 1.8, 1.87\}$, from left to right, accordingly with \eqref{frequencies-symmetric-case}. }
\end{figure}

The behavior of the energy velocity is represented in Fig.~\ref{plot-energy-velocity-symmetric-case-Drude-with-magnetic-current} in terms of $\omega/\omega_{0}$. As it happened in the antisymmetric case, see Sec.~\ref{section-antisymmetric-case}, the frequencies $\omega^{s}_{1, 2}$ define the region where occurs an absorption peak followed by an abrupt and strong reduction of the real part of the refractive indices. Also, we observe the existence of two lossy windows where the energy velocity is null. In details:

\begin{itemize}
	
	\item
		For $0<\omega < \omega^{s}_{1}$: there does not occur propagation, since $\mathrm{Re}[n_{\pm}]=0$ and the energy velocity is null, $V_{E}^{\pm}=0$.
	
	%
	\item
	For $\omega^{s}_{1} < \omega <\omega'$: propagation and absorption occur simultaneously and $V_{E}^{\pm} \neq 0$, since the real and imaginary pieces of $n_{\pm}$ are non-null, see Figs.~\ref{plot-n-parallel-symmetric-case-Drude-with-magnetic-current} and \ref{plot-n-antiparallel-symmetric-case-Drude-with-magnetic-current}. Whereupon the resonance frequency $\omega'$, the pieces $\mathrm{Re}[n_{\pm}]$ undergo a rapid deacreasing, implying a null energy velocity. 
\item
	For $\omega' < \omega < \omega^{s}_{2}$: only absorption occurs, since $\mathrm{Re}[n_{\pm}] =0$ and  $V_{E}^{\pm}=0$ (see Figs.~\ref{plot-n-antiparallel-symmetric-case-Drude-with-magnetic-current} and \ref{plot-energy-velocity-symmetric-case-Drude-with-magnetic-current}).
	
	\item Finally, for $\omega >\omega^{s}_{2}$: the propagation of both modes $n_{\pm}$ is restored, while the absorption effect diminishes rapidly to zero. The energy velocity increases monotonically to its asymptotic value.
\end{itemize}

To obtain the characteristic frequencies, we need to solve the following the sixth-order equation: 
\begin{eqnarray}
	&&
	\gamma^{2} \left( \omega^{4} - s \, \omega_{0}^{2} \right) +(\omega_{0}^{2}-\omega^{2}) \times
	\nonumber \\
	&&
	\times  \left[ \omega^{2} \left(\omega_{0}^{2}-\omega^{2}+\omega_{p}^{2} \right) - s  \left(\omega_{0}^{2}-\omega^{2}-\gamma^{2} \right) \right] =0
	\; , \hspace{0.5cm}
	\label{frequencies-symmetric-case}
\end{eqnarray}
with 
\begin{align}
	s=\frac{\mu}{16\epsilon_{0}} \, |{\bf{a}}\times {\bf{c}}|^{2} \cos^{2}\varphi \; ,
	\label{frequencies-symmetric-case-1}
\end{align}
in the place of the factor 
${\mu b^{2}\cos^{2}\theta}/{4\epsilon_{0}}$ in Eq. (\ref{frequencies-antisymmetric-case}). \eqref{frequencies-symmetric-case} provides the roots $\omega^{s}_{1, 2, 3}$, when $|{\bf{a}}\times {\bf{c}}| \cos\varphi \neq 0$, and $\omega^{\prime, \prime\prime}$ when $|{\bf{a}}\times {\bf{c}}| \cos\varphi =0$. We mention that $\omega^{s}_{3} \approx \omega'$, then we have depicted only $\omega'$ in the previous plots.

%

%

%

\section{Final Remarks}
\label{final}
In this paper, we have discussed the properties of wave propagation and absorption in a Drude-Lorentz dielectric in the presence of magnetic conductivity. The analysis was performed for three configurations of the magnetic conductivity tensor: isotropic, antisymmetric, and symmetric. For each case, we have obtained the dispersion relation and refractive indices as functions of the frequency and of the Drude-Lorentz parameters, with their properties properly scrutinized for all configurations. The involved refractive indices were examined in the range of low- and high-frequencies, being always compared to Drude's usual dielectric behavior. The group and the energy velocities were evaluated for all cases, being the latter one used as tool to examine signal propagation.
	
In lossy scenarios, as the group velocity becomes complex and loses physical interpretation, the signal propagation passes to be described by the energy velocity, which is real. In the high-frequency limit, one obtained
$\Re[n]=\sqrt{\mu\,\epsilon_{0}}$, while the energy velocity is reduced to
$V_{E}=1/\sqrt{\mu\epsilon_0}$ (for all the cases examined). The absorption goes to zero in this frequency regime, $\Im[n] \rightarrow 0$. Near the origin and at intermediary frequencies, however, the refractive indices behave in some different ways. 

The first case studied was the DD with isotropic magnetic conductivity, $\Sigma \neq 0$, examined in Sec.~\ref{section-isotropic-case}. The associated refractive indices show distinct features in relation to the conventional DD. Near the origin, it occurs $\Re[n_{+}] \rightarrow \infty$, approaching the behavior of a magnetized plasma index, while one has $\Re[n_{-}] \rightarrow 0$ (for $\omega \rightarrow 0$). These latter properties also manifest in a simple dielectric,  $\epsilon^{\prime}=cte$, $\epsilon^{\prime\prime}=\sigma/\omega$, with  $\Sigma=cte$ \cite{PedroMario}. The absorption window now has its length reduced by the magnetic conductivity. Moreover, the lossy interval begins to support signal propagation, $V_{E}^{\pm}\neq 0$, because  the refractive indices $n_{\pm}$ develop non-null real pieces in the range $\omega^{i}_{1} < \omega < \omega''$, in which the signal propagation is favoured. See Figs. \ref{plot-n-mais-isotropic-case-Drude-with-magnetic-current}, \ref{plot-n-menos-isotropic-case-Drude-with-magnetic-current} and \ref{plot-energy-velocity-isotropic-case-Drude-with-magnetic-current} for cheking details. Thus, the frequency window where $V_{E}=0$ is nearly supressed in the presence of the magnetic conductivity, $\Sigma \neq 0$.

For the antisymmetric and symmetric cases of Secs. \ref{section-antisymmetric-case} and \ref{section-symmetric-case}, respectively, we observe that the frequency window, where the energy velocity is null, is slightly ``enhanced'' due to the presence of the magnetic conductivity ($b\cos\theta \neq 0$, and $|{\bf{a}}\times {\bf{c}}| \cos\varphi\neq0$). Such a feature occurs because the non-null magnetic conductivity enlarges the frequency window where absorption effects can take place, see Figs.~\ref{plot-n-antisymmetric-parallel-case-Drude-with-magnetic-current} and \ref{plot-n-antisymmetric-antiparallel-case-Drude-with-magnetic-current} for the antisymmetric case, and Figs.~\ref{plot-n-parallel-symmetric-case-Drude-with-magnetic-current} and \ref{plot-n-antiparallel-symmetric-case-Drude-with-magnetic-current} for the symmetric case. In both configurations, the energy velocity becomes predominantly null for frequencies approximately smaller than $2\omega_{0}$, a window in which the signal propagation is severely constrained. See the plots of Fig. \ref{plot-energy-velocity-antisymmetric-case-Drude-with-magnetic-current} and Fig. \ref{plot-energy-velocity-symmetric-case-Drude-with-magnetic-current}. Such a behavior states a sharp difference in relation to the DD with isotropic conductivity of Sec.~\ref{section-isotropic-case}, which presents enhanced progapation in this window, as already mentioned.  The properties of the DD refractive indices and DD with magnetic conductivity are displayed in Table \ref{Drude-dieletric-modified-scenarios-comparison}.

As it is known in the literature, in non-absorbing media the energy velocity and group velocity are equal, $v_{g}=v_{E}$, equivalence that disappears in lossy media where it holds $v_{g}\neq v_{E}$. In the present work 
all the analyzed scenarios are endowed with absorption and  $\epsilon''=0$, which justifies the attainment of $v_{g}\neq v_{E}$ in all sections of the work
.

\begin{widetext}
  	\hspace{2cm}  \begin{table}[H]
  		\caption{Comparison of propagation properties of each modified scenario.}
  		\centering
  		\begin{tabular}{ C{3.9cm}   C{4cm}  C{3.5cm}   C{3cm}      C{2.8cm} }
  			\toprule
  			&   refractive indices at origin& absorption zone &   negative refraction & signal propagation    \\[0.6ex]
  			\colrule \\[0.6ex]
  			Drude dielectric (DD) &
  			$n \neq 0$, $n\in \mathbb{R}$   &         intermediary one, whereupon the resonance            &  no  & yes, except in the window $\omega^{i}_{1}<\omega <\omega''$ of Fig. \ref{plot-energy-velocity-isotropic-case-Drude-with-magnetic-current}
  			\\
  			[0.6ex]
  			\\
  			[0.6ex]
  			DD with $\Sigma$ &        $n_{+} \rightarrow \infty$  \quad \quad  \quad \quad  \quad $n_{-} \rightarrow 0$ \quad \quad \quad \quad  \quad $\Im[n_{\pm}] =0$         & one intermediary reduced window     &  \vspace{-.5cm}no \quad \quad \quad \quad  \quad  \quad \quad yes   & yes, favoured for  $\omega \lesssim  2\omega_{0}$ \\
  			[0.6ex]
  			\\
  			[0.6ex]
  			DD/antisymmetric $\sigma^{B}_{ij}$        &  $\Re[n_{\pm}] =0$ for  $\theta =0$ \quad \quad \quad $\Im[n] \rightarrow 0$ for $\theta=0$  \quad \quad $\Im[n] \rightarrow \infty$ for $\theta=\pi$ &    two zones: i) near the origin;  ii) whereupon the resonance (enlarged)  & no & severely constrained for $\omega \lesssim  2\omega_{0}$
  			\\[0.6ex]
  			\\
  			[0.6ex]
  			DD/symmetric $\sigma^{B}_{ij}$        & $\Re[n_{\pm}] =0$ \quad \quad \quad \quad $\Im[n_{+}, n_{-}] \rightarrow 0$ for $\varphi=0,\pi$ \quad $\Im[n_{-}, n_{+}] \rightarrow \infty$ for $\varphi=0,\pi$  \quad \quad     &    two zones:  i) near the origin;  ii) whereupon the resonance (enlarged) & no  &severely  constrained for $\omega \lesssim  2\omega_{0}$
  			\\[0.6ex]
  			\botrule
  		\end{tabular}
  		\label{Drude-dieletric-modified-scenarios-comparison}
  	\end{table}
  \end{widetext}

\begin{acknowledgments}
	
The authors express their gratitude to FAPEMA, CNPq, and CAPES (Brazilian research agencies) for their invaluable financial support. M.M.F. is supported by  CNPq/Produtividade 311220/2019-3 and CNPq/Universal/422527/2021-1. P.D.S.S is grateful to grant CNPq/PDJ 150584/23. Furthermore, we are indebted to CAPES/Finance Code 001 and FAPEMA/POS-GRAD-02575/21.

\end{acknowledgments}

\bigskip


\begin{thebibliography}{30}
%

\bibitem{Zangwill}
A.~Zangwill, \textit{Modern Electrodynamics}. New York (USA): Cambridge University Press, 2012.

\bibitem{Jackson} J. D. Jackson, {{\textit{{Classical Electrodynamics}}}}, 3rd edition. New York (USA): John Wiley \& Sons, 1999.

\bibitem{Landau} L. D. Landau and E. M. Lifshitz, \textit{Electrodynamics
of continuous media, Course of Theoretical Physics, Volume 8}, 2nd ed. (Pergamon Press, New York, 1984).

\bibitem{FeynmanQED}  R. P. Feynman, {QED: The Strange Theory of Light and Matter}, Princeton University Press (1985).


\bibitem{Drude1} Paul Drude, {\it Zur Elektronentheorie der Metalle}, \href{https://doi.org/10.1002/andp.19003060312}{Annalen der Physik, {\bf 306} (3): 566–613 (1900)}.


\bibitem{Drude2} Paul Drude, {\it Zur Elektronentheorie der Metalle; II. Teil. Galvanomagnetische und thermomagnetische Effecte}, \href{https://doi.org/10.1002/andp.19003081102}{Annalen der Physik. {\bf 308} (11): 369–402 (1900)}.


\bibitem{Ascrofit} Neil Ashcroft and N. David Mermin, {\it Solid State Physics}, New York: Holt, Rinehart and Winston (1976).


\bibitem{Fox} M. Fox, {Optical properties of Solids}, New York: Oxford University Press (2001).

\bibitem{Wooten} F. Wooten, {Optical properties of Solids}, New York: Academic Press (1972).

\bibitem{Kharzeev1} D. E. Kharzeev,  The chiral magnetic effect and anomaly induced transport, \href{https://doi.org/10.1016/j.ppnp.2014.01.002}{Prog. Part. Nucl. Phys. {\bf 75}, 133 (2014)}.
%


%
\bibitem{Kharzeev2} D. E. Kharzeev, J. Liao, S. A. Voloshin, and
G. Wang, Chiral magnetic and vortical effects in highenergy nuclear collisions – A status report, \href{https://doi.org/10.1016/j.ppnp.2016.01.001}{Prog. Part. Nucl. Phys. {\bf 88}, 1 (2016)}.


\bibitem{Kharzeev3} D. Kharzeev, K. Landsteiner, A. Schmitt and H.U. Yee, {\it Strongly Interacting Matter in
Magnetic Fields}, Lect. Notes Phys. {\bf 871} (Springer-Verlag,Berlin Heidelberg, 2013).


\bibitem{Dvornikov} M. Dvornikov and V.B. Semikoz,  Influence of the turbulent motion on the chiral magnetic effect in the early universe, \href{https://doi.org/10.1103/PhysRevD.95.043538}{Phys. Rev. D {\bf 95}, 043538 (2017)}.


\bibitem{Schober} J. Schober, A. Brandenburg and I. Rogachevskii,  Chiral fermion asymmetry in high-energy plasma simulations, \href{https://doi.org/10.1080/03091929.2019.1591393}{Geophys. Astrophys. Fluid Dynamics {\bf 114}, 106 (2020)}.


\bibitem{Dvornikov2} M. Dvornikov and V. B. Semikoz,  Instability of magnetic
fields in electroweak plasma driven by neutrino asymmetries, \href{https://doi.org/10.1088/1475-7516/2014/05/002}{JCAP {\bf 05}, 002 (2014)}.


\bibitem{Dvornikov3} M. Dvornikov,  Chiral magnetic effect in the presence of an external axial-vector
field, \href{https://doi.org/10.1103/PhysRevD.98.036016}{Phys. Rev. D {\bf 98}, 036016 (2018)}.


\bibitem{Dvornikov4} M. Dvornikov, Electric current induced by an external
magnetic field in the presence of electroweak matter, \href{https://doi.org/10.1051/epjconf/201819105008}{EPJ Web Conf. {\bf 191}, 05008 (2018)}.


\bibitem{Burkov} A. A. Burkov, Chiral anomaly and transport in Weyl metals, \href{https://doi.org/10.1088/0953-8984/27/11/113201}{J. Phys. Condens. Matter {\bf 27}, 113201 (2015)}.

\bibitem{Li} Q. Li, D. E. Kharzeev, C. Zhang, Y. Huang, I. Pletikosić, A. V. Fedorov, R. D. Zhong, J. A. Schneeloch, G. D. Gu, and T. Valla, Chiral magnetic effect in ZrTe5, \href{https://doi.org/10.1038/nphys3648}{Nat. Phys. {\bf{12}}, 550 (2016)}.
\bibitem{Xiaochun-Huang} X. Huang, L. Zhao, Y. Long, P. Wang, D. Chen, Z. Yang, H. Liang, M. Xue, H. Weng, Z. Fang, X. Dai, and G. Chen, Observation of the chiral-anomaly-induced negative magnetoresistance in 3D Weyl semimetal TaAs, \href{https://doi.org/10.1103/PhysRevX.5.031023}{Phys. Rev. X {\bf{5}}, 031023 (2015)}.
\bibitem{Barnes} E. Barnes, J. J. Heremans, and Djordje Minic, Electromagnetic Signatures of the Chiral Anomaly in Weyl Semimetals, \href{10.1103/PhysRevLett.117.217204}{Phys. Rev. Lett. \textbf{117}, 217204 (2016)}.

\bibitem{Pedro1} P. D. S. Silva, M. M. Ferreira Jr., M. Schreck, and L. F. Urrutia,  Magnetic-conductivity effects on electromagnetic propagation in
	dispersive matter, \href{https://journals.aps.org/prd/abstract/10.1103/PhysRevD.102.076001}{Phys. Rev. D {\bf{102}}, 076001 (2020)}.
	
	
\bibitem{Askne} J. Askne and B. Lind, Energy of Electromagnetic Waves in the Presence of Absorption and Dispersion, \href{https://doi.org/10.1103/PhysRevA.2.2335}{Phys. Rev. A {\bf 2}, 6 (1970)}.



\bibitem{Kaushik1} S.~Kaushik, D.E.~Kharzeev, and E.J.~Philip, Transverse chiral magnetic photocurrent induced by linearly polarized light in symmetric Weyl semimetals, \href{https://doi.org/10.1103/PhysRevResearch.2.042011}{Phys. Rev. Research {\bf{2}}, 042011(R) (2020)}.


\bibitem{ChenPRA}  P. Y. Chen, R. C. McPhedran, C. M. de Sterke, C. G. Poulton, A. A. Asatryan, L. C. Botten, and M. J. Steel, Group velocity in lossy periodic structured media, \href{https://doi.org/10.1103/PhysRevA.82.053825}{Phys. Rev. A \textbf{82}, 053825 (2010).}
	
	

\bibitem{Brillouin}  L. Brillouin, Wave Propagation and Group Velocity (Academic Press, New York, 1960).

\bibitem{Loudon} R. Loudon, The propagation of electromagnetic energy through an absorbing dielectric, \href{https://doi.org/10.1088/0305-4470/3/3/008}{J. Phys. A: Gen. Phys.{\bf{3}}, 233 (1970)}.
\bibitem{Loudon2} J. Neufeld, Wave propagation and group velocity in absorbing media,
\href{https://doi.org/10.1016/0375-9601(69)91036-6}{Phys. Lett. A Phys. {\bf{29}}, 68 (1969)}.
\bibitem{Sherman}
G. C. Sherman and K. E. Oughstun, Energy-velocity description of pulse propagation in absorbing, dispersive dielectrics, \href{https://doi.org/10.1364/JOSAB.12.000229}{J. Opt. Soc. Am. B \textbf{12}, 229-247 (1995)}.
\bibitem{Davidovich} M.V. Davidovich, Electromagnetic energy density and velocity in a medium with anomalous positive dispersion, \href{https://doi.org/10.1134/S106378500611023X}{Tech. Phys. Lett. \textbf{32}, 982–986 (2006).}
\bibitem{Ruppin}
R. Ruppin, Electromagnetic energy density in a dispersive and absorptive material, \href{https://doi.org/10.1016/S0375-9601(01)00838-6}{Phys. Lett. A {\bf{299}}, 309-312 (2002)}.
\bibitem{Nunes} F. D. Nunes, T. C. Vasconcelos, M. Bezerra, and J. Weiner, Electromagnetic energy density in dispersive
and dissipative media, \href{https://doi.org/10.1364/JOSAB.28.001544}{J. Opt. Soc. Am. B, {\bf 28} 6 (2011)}.

\bibitem{Moradi} A. Moradi and Pi‑G. Luan, Electromagnetic energy density
	in hyperbolic metamaterials, \href{https://doi.org/10.1038/s41598-022-14909-0}{Sci. Rep. {\bf 12}, 10760  (2022)}.
	\bibitem{Darinskii} A. N. Darinskii, Group and energy velocities of electromagnetic waves in bianisotropic superlattices, \href{https://doi.org/10.1103/PhysRevA.108.013510}{Phys. Rev. A {\bf 108}, 013510 (2023)}.
\bibitem{Cui} T. J. Cui and J. A. Kong, Time-domain electromagnetic energy in a frequency-dispersive left-handed medium,
		\href{https://doi.org/10.1103/PhysRevB.70.205106}{Phys. Rev. B \textbf{70}, 205106 (2004).}
\bibitem{Semchenko} I. Semchenko, A. Balmakou, S. Khakhomov, and S. Tretyakov, Stored and absorbed energy of fields in lossy chiral single-component metamaterials, \href{https://doi.org/10.1103/PhysRevB.97.014432}{Phys. Rev. B \textbf{97}, 014432 (2018).}	
\bibitem{Chen21} J. Chen, Y. Dai and Y. Xuanyuan, A possible way to experimentally examine validity of the expressions of dissipated energy density, \href{https://doi.org/10.1016/j.ijleo.2020.165756}{Optik {\bf 242}, 165756 (2021)}.
\bibitem{Chen22} J. Chen and J. She, Analysis on energy density difference between linearly and circularly polarized electromagnetic waves, \href{https://doi.org/10.1140/epjp/s13360-022-02697-5}{ Eur. Phys. J. Plus {\bf 137}, 502 (2022)}.
\bibitem{Gerasik} V. Gerasik and M. Stastna, Complex group velocity and energy transport in absorbing media, \href{https://doi.org/10.1103/PhysRevE.81.056602}{Phys. Rev. E {\bf{81}}, 056602 (2010)}.


\bibitem{PedroMario} P. D. S. Silva, M. J. Neves and M. M. Ferreira Jr., Group velocity and energy propagation in a dielectric medium supporting magnetic current, \href{https://doi.org/10.48550/arXiv.2305.08153}{arXiv:2305.08153v1(2023)}.




\bibitem{Filipe} F. S. Ribeiro, P. D. S. Silva, and M. M. Ferreira Jr., Cold plasma modes in the chiral Maxwell-Carroll-Field-Jackiw electrodynamics, \href{https://doi.org/10.1103/PhysRevD.107.096018}{Phys. Rev. D {\bf 107}, 096018 (2023)}.


\bibitem{Hosur}
P. Hosur, and X-L. Qi, Tunable circular dichroism due to the chiral anomaly in Weyl semimetals, \href{https://doi.org/10.1103/PhysRevB.91.081106}{Phys. Rev. B \textbf{91}, 081106(R) (2015)}.


\bibitem{Nieto-Vesperinas}
M. Nieto-Vesperinas, Optical theorem for the conservation of electromagnetic helicity: Significance for molecular energy transfer and enantiomeric discrimination by circular dichroism, \href{https://doi.org/10.1103/PhysRevA.92.023813} {Phys. Rev. A \textbf{92}, 023813 (2015)}.


\bibitem{Tang} Y. Tang and A. E. Cohen, Enhanced Enantioselectivity in Excitation of Chiral Molecules by Superchiral Light,   \href{https://doi.org/10.1126/science.1202817}{Science \textbf{332}, 333 (2011)}.

\bibitem{Amin}
M. Amin, O. Siddiqui, and M. Farhat, Linear and circular dichroism in graphene-based reflectors for polarization control, \href{https://doi.org/10.1103/PhysRevApplied.13.024046}{Phys. Rev. Applied \textbf{13}, 024046 (2020)}.
\bibitem{Shibata} J.~Shibata, A.~Takeuchi, H.~Kohno, and G.~Tatara, Theory of electromagnetic wave propagation in ferromagnetic Rashba conductor, \href{https://aip.scitation.org/doi/10.1063/1.5011130}{J. App. Phys. \textbf{123}, 063902 (2018)}.


%
\end{thebibliography}
\end{document}